\documentclass[12pt]{amsart}
\usepackage{geometry}

\usepackage{setspace}
\usepackage{amssymb}
\usepackage{times}
\usepackage{fullpage}
\usepackage{color}
\usepackage{graphicx}
\usepackage{bm}
\usepackage{textcomp}

\makeatletter
\def\specialsection{\@startsection{section}{1}%
  \z@{\linespacing\@plus\linespacing}{.5\linespacing}%
  {\normalfont}}
\def\section{\@startsection{section}{1}%
  \z@{.7\linespacing\@plus\linespacing}{.5\linespacing}%
  {\normalfont\scshape}}
\makeatother

\title{Emerging superconductivity with broken time reversal symmetry inside a superconducting $s$-wave state.}

\author{V.\,G\MakeLowercase{rinenko}$^{1,2,*}$ \and 
R.\,S\MakeLowercase{arkar}$^{1}$ \and K.\,K\MakeLowercase{ihou}$^{3}$ \and C.H.\,L\MakeLowercase{ee}$^{3}$ \and I.\,M\MakeLowercase{orozov}$^{2,4}$ \and S.\,A\MakeLowercase{swartham}$^{2}$ \and B.\,B\MakeLowercase{\"uchner}$^{2}$ \and P.\,C\MakeLowercase{hekhonin}$^{1,2}$ \and   W.\,S\MakeLowercase{krotzki}$^{1}$ \and K.\,N\MakeLowercase{enkov}$^{2}$ \and R.\,H\MakeLowercase{\"uhne}$^{2}$ \and K.\,N\MakeLowercase{ielsch}$^{2}$ \and D.V.\,E\MakeLowercase{fremov}$^{2}$ \and S.-L.\,D\MakeLowercase{rechsler}$^{2}$ \and V.L.\,V\MakeLowercase{adimov}$^{5}$ \and M.A.\,S\MakeLowercase{ilaev}$^{6}$ \and P.\,V\MakeLowercase{olkov}$^{7,8}$ \and I.\,E\MakeLowercase{remin}$^{7}$ \and H.\,L\MakeLowercase{uetkens}$^{9}$ \and H.-H.\,K\MakeLowercase{lauss}$^{1}$  
}

\begin{document}
\maketitle
\doublespacing

{\footnotesize
\noindent
1. Institute for Solid State and Materials Physics, TU Dresden, D-01069, Dresden, Germany\\
2. IFW Dresden, 01069, Dresden, Germany\\
3. National Institute of Advanced Industrial Science and Technology (AIST), Tsukuba, Ibaraki 305-8560 Japan\\
4. Lomonosov Moscow State University, Leninskie Gory, Moscow, 119991, Russian Federation\\
5. Institute for Physics of Microstructures, RAN, Nizhny Novgorod, GSP-105, Russia\\
6. Department of Physics and Nanoscience Center, University of Jyv\"askyl\"a, FI-40014, Finland\\
7. Institut f\"ur Theoretische Physik III, Ruhr-Universitat Bochum, 44801 Bochum, Germany\\
8. Department of Physics and Astronomy, Center for Materials Theory, Rutgers University, Piscataway, New Jersey 08854, USA\\
9. Laboratory for Muon Spin Spectroscopy, PSI, CH-5232 Villigen PSI, Switzerland\\

} 

\section*{Abstract}
\noindent{\bf
{In general, magnetism and superconductivity are antagonistic to each other. However, there are several families of superconductors, in which superconductivity may coexist with magnetism, and only a few examples are known, when superconductivity itself induces spontaneous magnetism. The most known compounds are Sr$_2$RuO$_4$ and some noncentrosymmetric superconductors. Here, we report the finding of a narrow dome of a novel $s+is'$ superconducting (SC) phase with broken time-reversal symmetry (BTRS) inside the broad $s$-wave SC region of the centrosymmetric multiband superconductor Ba$_{\rm 1-x}$K$_{\rm x}$Fe$_2$As$_2$ ($0.7 \lesssim x \lesssim 0.85$). We observe spontaneous magnetic fields inside this dome using the muon spin relaxation ($\mu$SR) technique. Furthermore, our detailed specific heat study reveals that the BTRS dome appears very close to a change in the topology of the Fermi surface (Lifshitz transition). With this, we experimentally demonstrate the emergence of a novel quantum state due to 
topological changes of the electronic system.}}

\begin{figure}
\centering
\includegraphics[width=14cm]{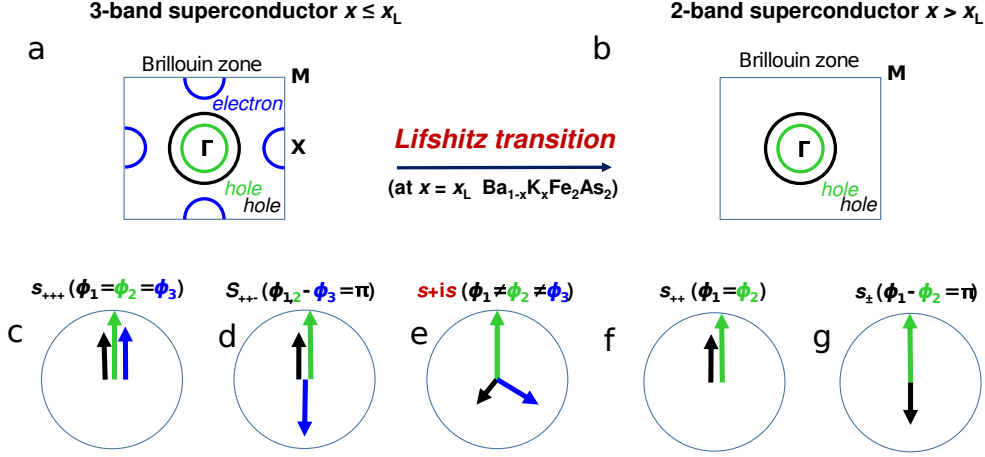}
\caption{{\bf Multiband superconductivity} Schematic illustration of the change of the topology of the Fermi surface at the Lifshitz transition in the Ba$_{\rm 1-x}$K$_{\rm x}$Fe$_2$As$_2$ system with possible $s$-wave superconducting states \cite{note_Fig1}.  a) The Brillouin zone with two hole and one electron Fermi pockets (optimal doping) and b) the Brillouin zone with two hole Fermi pockets (high doping level). Bottom panels c) - g): possible $s$-wave pairing states in a clean limit. The relative phase $\phi$ of the superconducting order parameter components is shown by the direction of the arrows inside the circles and the magnitude by the length. A frustrated pairing $s+is$ state with arbitrary phase shifts between the components of the order parameter in a clean limit is possible in the three-band case.} 
\label{Fig:1}
\end{figure}

The complexity of the Fermi surface may result in a competition between different SC pairing symmetries in multiband systems \cite{Lee2009, Ng2009, Tanaka2010, Stanev2010, Carlstrom2011, Hu2011, Khodas2012, Maiti2013, Garaud2014, Platt2015, Maiti2015, Garaud2016, Lin2016, Boker2017, Yerin2017, Vadimov2018}. For a two-band $s$-wave superconductor, the interband phase difference is either 0 or $\pi$ in the ground state (Fig.\ \ref{Fig:1}). However, for a superconductor with more than two bands, repulsive interband interactions might result in a complex superconducting (SC) order parameter with the interband phase difference being neither 0 nor $\pi$. Such a complex $s+id$ or $s+is'$ ground state breaks time-reversal symmetry (BTRS) and can be considered as a result of the competition between $s$ and $d$ or different $s$- wave components. These states with a frustrated phase of the order parameter  (Fig.\ \ref{Fig:1}e) \cite{Stanev2010,Chubukov2015} are {\it qualitatively} different from the previously studied systems with chiral pairing states such as Sr$_2$RuO$_4$ \cite{Luke1998} or with a generic triplet component like noncentrosymmetric superconductors \cite{Singh2014,Biswas2013}. 

In all known multiband superconductors, the phase difference of the SC order parameter between different bands was observed to be either 0 or $\pi$. For example, most iron based superconductors (FeSCs) are in close proximity to a SDW phase, in which the interaction between fermions in hole ($h$) and electron ($el$) like Fermi pockets is strongly enhanced giving rise to an $s_{\rm \pm}$ state \cite{Mazin2010, Kuroki2008}. Therefore, frustrated superconductivity might be expected in FeSCs, which are quite far from the SDW phase \cite{Ahn2014}. 
It has been argued that in the Ba$_{\rm 1-x}$K$_{\rm x}$Fe$_2$As$_2$ system the change of the SC order parameter with doping is caused by topological changes of the Fermi surface (FS), i.e. at a Lifshitz transition (schematically shown in Fig.\ \ref{Fig:1}) \cite{Maiti2013, Boker2017}. According to angle-resolved photoemission spectroscopy (ARPES) at optimal doping, the FS consist of $h$ and $el$ like pockets \cite{Ding2008, Kordyuk2014}. This band structure favors an $s_{\rm \pm}$ state with a gap function having different signs ($\pi$ phase shift) on the $h$ and $el$ pockets (Fig.\ \ref{Fig:1}d). With further K-doping the $el$ pockets disappear and additional propeller like $h$ pockets appear at the Brillouin zone corner \cite{Xu2013,Malaeb2012} (not shown in Fig.\ \ref{Fig:1} for simplicity). These changes suppress the $h$ - $el$ interpocket interaction resulting in a change of the  $s_{\rm \pm}$ SC gap symmetry to another $s$ or a $d$ - wave state. As a consequence the frustration of the order parameter due to competing interband interactions may lead to phase shifts different from 0 and $\pi$ (Fig.\ \ref{Fig:1}e). It has been predicted that close to the Lifshitz transition, incipient $el$ bands still contribute to the SC pairing until the distance from the top of these pockets to the Fermi level are comparable with the SC gap size \cite{Boker2017}.  Therefore, it is expected that this frustrated BTRS SC state appears in the Ba$_{\rm 1-x}$K$_{\rm x}$Fe$_2$As$_2$ system at a doping level, at which the $el$ bands just sank below the Fermi level. 

It was shown that in the presence of inhomogeneities $s+is'$ and $s+id$ phases can generate spontaneous currents and related magnetic fields \cite{Garaud2014, Lin2016, Vadimov2018}. These spontaneous magnetic fields in the SC state can be observed experimentally. In our first muon spin relaxation ($\mu$SR)  experiments, we detected an enhancement of the zero field (ZF) muon spin depolarization rate below the SC transition temperature $T_{\rm c}$ in heavy-ion irradiated Ba$_{\rm 1-x}$K$_{\rm x}$Fe$_2$As$_2$ single crystals with $x \approx$ 0.73 \cite{Grinenko2017}. 
The observed behavior indicates a BTRS in the SC state. However, the nature of the BTRS state remains elusive so far. In particular, a very strong inhomogeneity might result at low temperatures in a BTRS state even in a two-band $s_{\pm}$ superconductor \cite{Bobkov2011}. Here, we performed systematic investigations of high-quality Ba$_{\rm 1-x}$K$_{\rm x}$Fe$_2$As$_2$ single crystals to elucidate the fundamental mechanism responsible for the spontaneous magnetic fields in the SC state. By combined $\mu$SR and specific heat studies we show that the time reversal symmetry is indeed spontaneously broken inside a singlet $s$-wave SC region in clean Ba$_{\rm 1-x}$K$_{\rm x}$Fe$_2$As$_2$ single crystals. We also define the size of the BTRS dome on the phase diagram, the symmetry of the SC order parameter in the BTRS state, and the relationship of this novel SC state with topological changes of the Fermi surface.

\begin{figure}
	\centering
		\includegraphics[width=14cm]{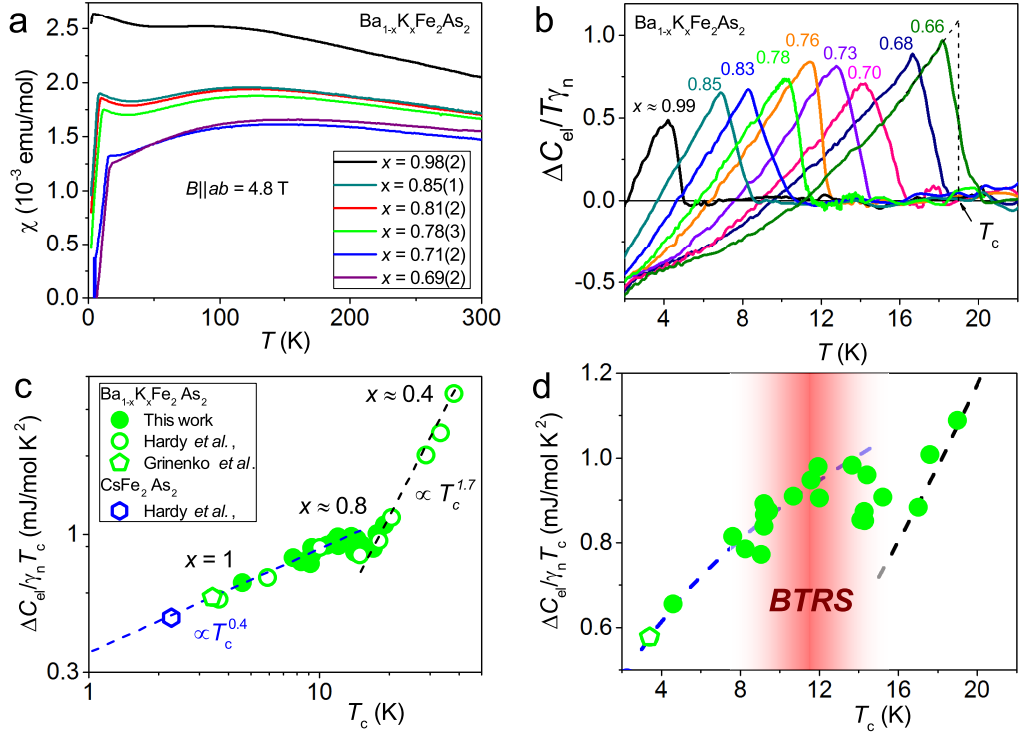}
		\caption{{\bf Thermodynamic properties} a) Temperature dependence of the molar susceptibility $\chi$ of the Ba$_{\rm 1-x}$K$_{\rm x}$Fe$_2$As$_2$ single crystal stacks used in the $\mu$SR experiments. b)
 Temperature dependence of the electronic specific heat of Ba$_{\rm 1-x}$K$_{\rm x}$Fe$_2$As$_2$ single crystals with various doping levels. c) Double logarithmic plot of the normalized specific heat jump $\Delta C_{\rm el}/\gamma_{\rm n}T$ at $T_{\rm c}$ versus $T_{\rm c}$ obtained from the entropy construction as shown in Fig.\ \ref{Fig:2}b (dotted line). Closed symbols - data from this work, open symbols - data taken from literature \cite{Grinenko2014,Hardy2016}. d) The $\Delta C_{\rm el}/\gamma_{\rm n}T$ close to the region with the BTRS state observed in the $\mu$SR experiments.}
\label{Fig:2}
\end{figure}

The SC and normal state physical properties of Ba$_{\rm 1-x}$K$_{\rm x}$Fe$_2$As$_2$ single crystals were characterized by DC susceptibility and specific heat measurements. The magnetic susceptibility in the normal state of the samples used in the $\mu$SR experiments is shown in Fig.\ref{Fig:2}a. The systematic increase of the susceptibility with K doping is consistent with the reported data \cite{Liu2014}. The increase of the susceptibility correlates with an enhancement of the electronic specific heat in the normal state (see below). A pronounced narrow anomaly in the specific heat at $T_{\rm c}$ indicates the high quality of the crystals (see Fig.\ \ref{Fig:2}b). X-ray and transmission electron microscopy (TEM) did not reveal any secondary phase or extended crystalline defects (see Fig.\ S1 in the Supplementary material (SM)). For the $\mu$SR experiments we selected samples with doping levels of $x$ = 0.69(2) , 0.71(2), 0.78(3), 0.81(2), 0.85(1) and 0.98(2). Additional information on the properties of the samples and the experimental design is given in the Method section and the SM. 

\begin{figure}
	\centering
		\includegraphics[width=12.5cm]{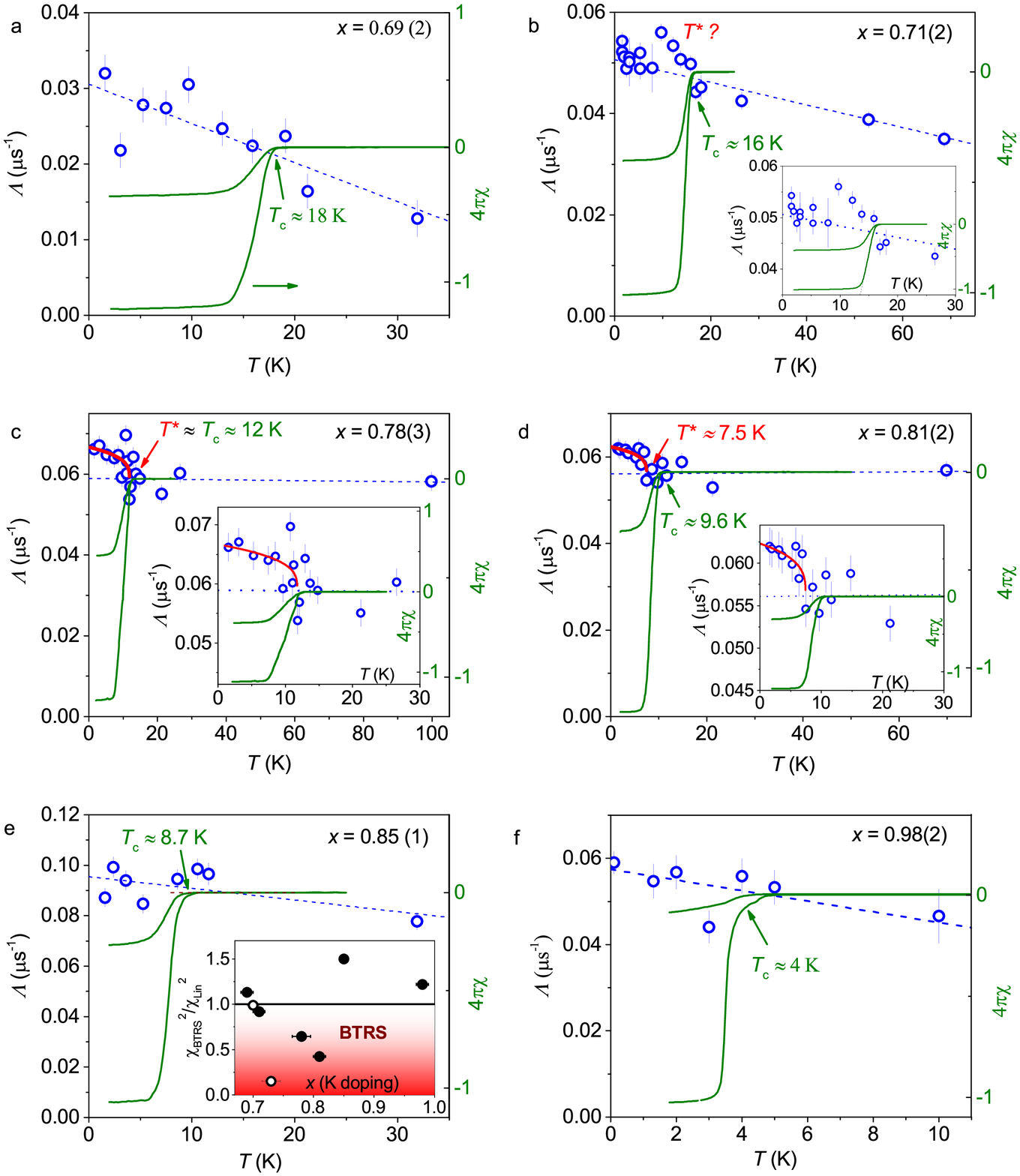}
		\caption{{\bf Zero field $\mu$SR} 
(Left axis) Temperature dependence of the muon spin depolarization rate $\Lambda$ for the muon spin component $P_{\rm \mu} \parallel c$ axis for Ba$_{\rm 1-x}$K$_{\rm x}$Fe$_2$As$_2$ samples with different K doping. The blue curves are linear fits at $T \geq T^*$. The red curves are fits at $T \leq T^*$. (Right axis) Temperature
dependence of the volume susceptibility measured in a low magnetic-field $B \parallel ab$ = 5 G applied after cooling in zero field, subsequent warming in the field (ZFC), and cooling again in the same field (FC) for samples with, a) $x$ = 0.69(2), b) $x$ = 0.71(2), c) $x$ = 0.78(3), d) $x$ = 0.81(2), e) $x$ = 0.85(1), and d)  $x$ = 0.98(2). Insets of Figs.\ \ref{Fig:3}b, \ref{Fig:3}c, and \ref{Fig:3}d: zoom in the data at low temperatures, and the inset of Fig.\ \ref{Fig:3}e shows the ratio $\chi_{\rm BTRS}^2/\chi_{\rm Lin}^2$ for the fits with and without including the $\Delta\Lambda$. Closed symbols - data from this work, open symbols - data for the samples with $x$ =0.7(2) and $x$ = 0.73(2) taken from Ref. \cite{Grinenko2017}.}
\label{Fig:3}
\end{figure}

The temperature dependencies of the ZF muon spin depolarization rate $\Lambda$ together with the magnetic susceptibility in the SC state are shown in Fig. \ref{Fig:3}, the raw data together with details of the analysis are summarized in  Section 2 of the SM. The small normal state $\Lambda \lesssim$  0.1 $\mu$s$^{-1}$ is temperature and doping dependent. The extrapolated depolarization rate $\Lambda_0$ to $T$ = 0 monotonously increases with the doping following the behavior of the normal state bulk susceptibility (see Fig.\ \ref{FigS_ZF_relax_t0} in the SM). The increase of $\Lambda_0$ with K doping is in line with the behavior of the normal state spin-lattice relaxation rate $1/T_1$ and the NMR Knight shift \cite{Hirano2012} anticipating the proximity of KFe$_2$As$_2$ to an yet unknown quantum critical point (QCP) \cite{Eilers2016, Drechsler2017, Drechsler2018}. This indicates that the depolarization rate $\Lambda$ reflects intrinsic electronic properties of the samples.  

For the samples with $0.71 \lesssim x \lesssim 0.81$ we observe a systematic increase of $\Lambda$ in the SC state below a characteristic temperature $T^*$. The effect $\Delta\Lambda \sim  0.01 \mu s^{-1}$ is relatively small, but it is visible in the time dependence of the asymmetry measured for different temperatures (see Figs.\ \ref{FigS_muSR_1_12K}, \ref{FigS_muSR_2_12K}, \ref{FigS_muSR_1_10K}, and \ref{FigS_muSR_2_10K} in the SM). 
To determine the $T^*$ values we fitted the temperature dependencies of $\Lambda$ using a power law in reduced temperature $\Lambda = \Delta\Lambda(1-T/T^{*})^{\beta}+\Lambda_0 + b_{\Lambda}T$, 
where $\Delta\Lambda$ is the additional depolarization rate below $T^*$, $\beta$ is an exponent, and $\Lambda_0$ and $b_{\Lambda}$ describe a linear dependence of the $\Lambda$ in the normal state. We found that the quality of the fits, characterized by $\chi^2$, is improved after including the $\Delta\Lambda$ term for the samples with $0.71 \lesssim x \lesssim 0.81$, only. For other doping levels the linear fits result in smaller $\chi^2$ values. Doping dependence of 
the ratio $\chi_{\rm BTRS}^2/\chi_{\rm Lin}^2$ for the fits with and without including the 
$\Delta\Lambda$ term is shown in the inset to Fig.\ref{Fig:3}e. Accordingly, the samples with $\chi_{\rm BTRS}^2/\chi_{\rm Lin}^2 < 1$ we assigned to have the $\Delta\Lambda$ term  below  
$T^*$, wheres for other doping levels a change with respect to the normal state behavior is not observed. The best fits for the samples with well defined $T^*$  were obtained for $\beta \simeq 0.3$, hence $\beta = 0.3$ was fixed at the analysis to reduce the number of the fitting parameters. The $T^*$ values are summarized in Fig.\ \ref{Fig:5}. Note,
for the sample with $x = 0.71 (2)$ we cannot exclude a weak enhancement of $\Lambda$ in the SC state  (see also Figs.\ \ref{FigS_muSR_1_16K}, \ref{FigS_muSR_2_16K} in the SM). However, a transition temperature $T^*$ could not be unambiguously assigned. For this reason we did not include a point for this doping level in the phase diagram shown in Fig.\ \ref{Fig:5}. The enhancement of the depolarization rate is attributed to the appearance of spontaneous magnetic fields in the SC state at the BTRS transition temperature $T^*$. The doping range, in which the BTRS is observed, and the magnitude of $\Delta\Lambda$ is consistent with our previous observations for the heavy-ion-irradiated sample with $x \sim 0.73$ \cite{Grinenko2017}. However, a quantitative comparison is complicated due to the sensitivity of the local fields to the strength and orientation of the defects \cite{Lin2016, Vadimov2018}.

\begin{figure}
	\centering
		\includegraphics[width=14cm]{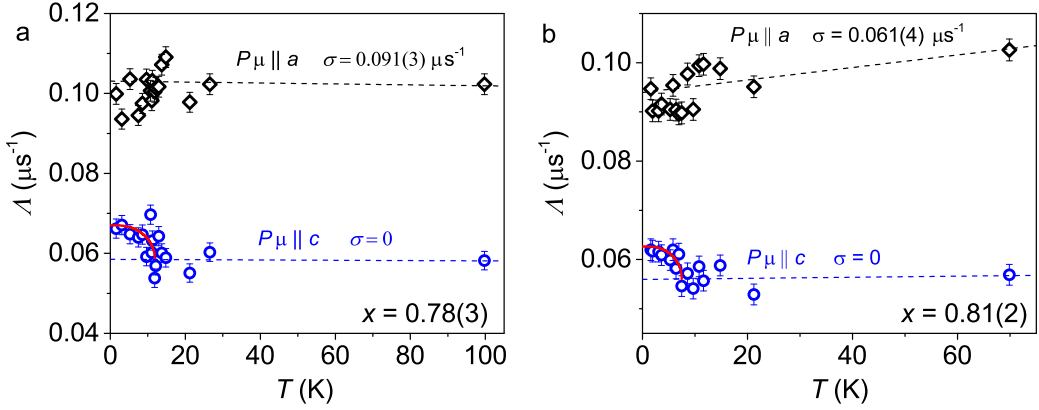}
		\caption{{\bf Anisotropy of the muon spin depolarization rate} Temperature dependence of the muon spin depolarization rate $\Lambda$ for two Ba$_{\rm 1-x}$K$_{\rm x}$Fe$_2$As$_2$ samples with BTRS state at low temperatures measured for the muon spin polarization components $P_{\rm \mu} \parallel c$  and $P_{\rm \mu} \parallel a$: a) sample with $x$ = 0.78(3), and b) sample with $x$ = 0.81(2). The dashed curves are linear fits at $T \geq T^*$, while the red curves are fits at $T \leq T^*$.}
\label{Fig:4}
\end{figure}

\begin{figure}
	\centering
		\includegraphics[width=14cm]{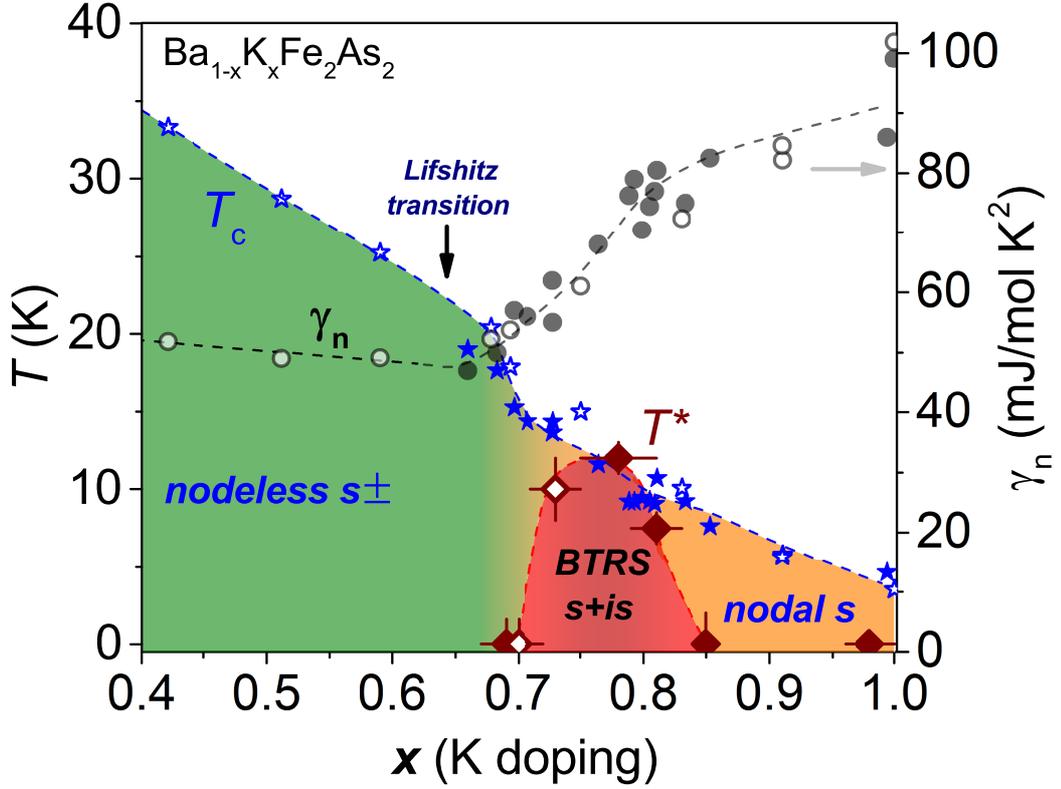}
		\caption{{\bf Phase diagram} (Left axis) Doping dependence of the 
SC transition temperature $T_{\rm c}$ and the onset temperature $T^*$ of the appearance of spontaneous magnetic fields in the superconducting state for the Ba$_{\rm 1-x}$K$_{\rm x}$Fe$_2$As$_2$ system. (Right axis) Doping dependence of the Sommerfeld coefficient in the electronic specific heat $\gamma_{\rm n}$. Closed symbols - data from this work, open symbols - data taken from  literature \cite{Grinenko2017,Hardy2016,Grinenko2014}. The doping range of the nodal superconductivity is shown according to Ref.\cite{Cho2018}. According to our specific heat study combined with the literature data \cite{Xu2013,Malaeb2012, Liu2014, footnote2, Hodovanets2014} the Lifshitz transition of the the electron pocket at the M points of the BZ occurs at $x_{\rm L} \simeq 0.55 - 0.7$.}
\label{Fig:5}
\end{figure}

In Fig.\ \ref{Fig:4} the temperature dependence of $\Lambda$ is shown within a broad temperature range  for two samples exhibiting a BTRS state. We present the data for two muon spin polarization components $P_{\rm \mu}\parallel c$ and $P_{\rm \mu}\parallel a$ measured along the crystallographic $c$ and $a$-axes (the data for other samples are given in Section 2 in the SM). In the normal state, $\Lambda$ is moderately anisotropic $\frac {\Lambda_{P_{\rm \mu}\parallel a}} {\Lambda_{P_{\rm \mu}\parallel c}} \sim 1.7$ and exhibits a nearly linear  temperature dependence. We could clearly detect the enhancement of the muon depolarization rate for $P_{\rm \mu}\parallel c$, whereas for $P_{\rm \mu} \parallel a$ such an enhancement is not observed.  Moreover, for $x = 0.78 (3)$ even a slight suppression of $\Lambda$ cannot be excluded below $T^*$ for  $P_{\rm \mu} \parallel a$. These data indicate that the internal fields in the BTRS state have a preferred orientation within the $ab$-plane \cite{footnote1}. 

To understand this anisotropy of the internal field strength, we modelled the dependence of the spontaneous magnetic fields  by adopting a local change of the superconducting coupling constants due to the sample inhomogeneities (see Fig.\ \ref{FigS6} the SM) using the approach proposed in Ref.\ \cite{Vadimov2018}. The sample inhomogeneities lead to the variation of the effective superconducting coupling constant $\lambda \approx \lambda_0 \pm \delta\lambda$ that parametrizes the interband pairing, where $\lambda_0$ is the coupling constant without inhomogeneities and $\delta\lambda$ describes the variation of the coupling constant due to them. We found that in a broad range of $\delta\lambda$ for an anisotropic $s+is'$ state the spontaneous magnetic fields are polarized mainly within the $ab$ crystal plane, while for an $s+id$ state their $ab$-plane and $c$-axis components would be of the same order \cite{Vadimov2018}. Thus, the observed strong anisotropy of the spontaneous internal fields can be understood as an indication for an anisotropic $s+is'$ SC state.

The magnitude of the additional depolarization rate in the BTRS state $\Delta\Lambda$ corresponds to an average internal field of $<B_{\rm int}> = \Delta\Lambda/\gamma_{\rm \mu} \sim$ 0.1 Oe, where $\gamma_{\rm \mu}$ = 0.085 $\mu$s$^{-1}$G$^{-1}$ is the gyromagnetic ratio of the muon. Such weak fields can be explained by the same model assuming a small variation of the SC coupling constants with $\pm\delta\lambda$ of a few percent (Fig.\ \ref{FigS6} in the SM). We attribute these weak defects to inhomogeneities in the doping level since no extended crystalline defects or impurity phases were found in our TEM investigations. In addition, the analysis of the electrical resistivity data (see Section 4 in the SM) indicates that the amount of disorder is nearly doping independent in the range of $x = 0.65 - 0.85$. Experimentally, the strength of these inhomogeneities $\delta\lambda$ can be estimated from the SC transition width measured by the specific heat. Assuming $\Delta T_{\rm c} =$ 1 - 2 K and using the BCS expression for $T_{\rm c} \propto exp(-1/\lambda)$ we get that $\delta\lambda$ varies within $\pm 5 \%$, which is in a reasonable agreement with the expected theoretical value (see the SM).

The evidence for a change of the SC order parameter with doping is provided by the $T_{\rm c}$ dependence of the specific heat jump $\Delta C_{\rm el}/\gamma_{\rm n}T$ at $T_{\rm c}$ (Fig.\ \ref{Fig:2}c). For Ba$_{\rm 1-x}$K$_{\rm x}$Fe$_2$As$_2$ single crystals with a doping level $x \lesssim 0.7$ $\Delta C_{\rm el}/\gamma_{\rm n}T_{\rm c}$  follows approximately the well-known BNC scaling behavior $\Delta C_{\rm el}/\gamma_{\rm n}T_{\rm c} \propto T_{\rm c}^\alpha$ with $\alpha \approx 2$ \cite{Budko2009}, which is considered to be a consequence of the nodeless multiband $s_{\rm \pm}$ superconductivity in iron pnictides \cite{Bang2017}. For the doping level $x \gtrsim 0.8$, $\Delta C_{\rm el}/\gamma_{\rm n}T_{\rm c} \propto T_{\rm c}^\alpha$ results in another scaling curve with $\alpha \approx 0.4$. This exponent value is also above the single-band BCS value $\alpha = 0$ ($\Delta C_{\rm el}/\gamma_{\rm n}T_{\rm c} = const$) and it can be obtained numerically within some theoretical models for an $s_\pm$ superconductor in the clean limit \cite{Bang2017}. The nodal gap structure for the Ba$_{\rm 1-x}$K$_{\rm x}$Fe$_2$As$_2$ system at this high doping level \cite{Ota2014, Watanabe2014, Cho2018} excludes a conventional $s_{\rm ++}$ - wave gap. Therefore, we expect an unconventional  superconductivity. It might be a different kind of sign change $s$-wave state. The simplest theoretical proposal (ignoring nodes) for such an $s'_\pm$ state is that the order parameter changes its sign between the two inner $h$ Fermi pockets (see Fig.\ \ref{Fig:1}g for the illustration) \cite{Maiti2013, Boker2017}. In the intermediate doping regime $0.7 \lesssim x \lesssim 0.8$, $\Delta C_{\rm el}/\gamma_{\rm n}T_{\rm c}$ behaves unusually by exhibiting a local maximum at the BTRS transition temperature $T^* \sim T_{\rm c}$ (Fig.\ \ref{Fig:2}d). The negative exponent $\alpha < 0$ for the intermediate region is striking, especially in view of the monotonously decreasing $T_{\rm c}$-value. The observed crossover between two distinct scaling behaviors in a narrow doping range indicates an essential modification of the SC pairing interactions. 

Evidence for a peculiar SC gap structure inside the BTRS phase can be found also in the high-resolution ARPES measurements \cite{Ota2014}. In the BTRS state, the nodes exist only on the outer $h$ Fermi surface centered at the $\Gamma$ point. Whereas at higher doping, outside the BTRS dome, additional nodes appear on the middle $h$ Fermi surface.     
Therefore, our specific heat data together with available ARPES data demonstrate that the BTRS state observed in the $\mu$SR measurements exists in an intermediate region of the phase diagram between two different kinds of unconventional $s$- wave SC states. 

To understand the microscopic reasons for the change of the SC order parameter symmetry, we performed detailed investigations of the electronic specific heat in the normal state (i.e. evaluating the Sommerfeld coefficient $\gamma_{\rm n}$). As in the case of the specific heat jump at $T_{\rm c}$, three different regions were found in the phase diagram shown in Fig.\ \ref{Fig:5}: i) Above optimal doping up to  $x \sim 0.65$, $\gamma_{\rm n}$ slightly reduces with doping in qualitative agreement with a linear suppression of $T_{\rm c}$. The reduction of $T_{\rm c}$ and $\gamma_{\rm n}$ is attributed to a weakening of the interband interactions due to a gradual reduction of the size of the $el$ Fermi pockets increasing the doping \cite{Xu2013, Malaeb2012, footnote2}. In this part of the phase diagram, the SC gap symmetry is $s_\pm$ with different signs of the order parameter on the $el$ and $h$ Fermi pockets (Fig.\ \ref{Fig:1}d). 

ii) For samples with a doping level $x \gtrsim 0.65$, $\gamma_{\rm n}$ steeply increases with doping. In contrast, $T_{\rm c}$ sharply reduces with the doping above $x \approx 0.7$ but flattens after the crossing of the BTRS-dome. In the BTRS state, the experimental data are consistent with an $s+is'$ frustrated SC state having a ground state interband phase difference of neither 0 nor $\pi$ (Fig.\ \ref{Fig:1}e). 
The steep increase of $\gamma_{\rm n}$ at $x \sim 0.65$ we related to the  Lifshitz transition observed in the ARPES data \cite{Xu2013,Malaeb2012, footnote2} at a similar doping level. The data in Ref.\ \cite{Malaeb2012} were obtained on the crystals prepared in the same way as our samples. In this case, the Lifshitz transition, related to the shift of the electron pockets at the M point of the Brillouin zone (BZ) above the Fermi level, was observed for the doping levels in between $x \sim 0.5$ ($T_{\rm c}$ = 31 - 35 K) and $x \sim 0.6$ ($T_{\rm c}$ = 23 - 27 K). According to these $T_{\rm c}$ values, in our phase diagram (Fig.\ref{Fig:5}) this corresponds to the doping levels $x$ = 0.4 - 0.5 and  $x$ = 0.55 - 0.63, correspondingly. In Ref.\ \cite{Xu2013} this Lifshitz transition  was observed for the doping levels in between 
$x = 0.7$ ($T_{\rm c} $ = 22 K) and $x = 0.9$ ($T_{\rm c}$ = 9 K). In our phase diagram this corresponds to the doping levels $x \approx$ 0.64 and $x \approx$ 0.83, correspondingly. Thus, we concluded that there is a reasonably good correspondence of the kink in the doping dependence of the Sommerefeld coefficient at $x \sim 0.65$ with the Lifshitz transition of the electron pocket. We also found a consistency between our data and transport measurements. For example, the minimum in the thermoelectric power at $x \sim$ 0.55  was attributed to the Lifshitz transition of the electronic Fermi surface \cite{Hodovanets2014}. According to the $c$-axis value, there is nearly exact correspondence between doping levels from this paper and our phase diagram. Finally, the Hall measurements strongly suggest that the the electron pocket at the M points contributes to the Hall coefficient up to the doping level $x \sim 0.65 - 0.8$ ($T_{\rm c} \sim$ 24 - 16 K) \cite{Liu2014}. According to $T_{\rm c}$ this doping level corresponds to $x \sim$ 0.6 - 0.7 in our phase diagram. In summary, our detailed specific heat study combined with the literature data \cite{Xu2013,Malaeb2012, Liu2014, footnote2, Hodovanets2014} allows to locate in the phase diagram the position of the Lifshitz transition of the the electron pocket at the M points at $x_{\rm L} \simeq 0.55 - 0.7$ (Fig.\ \ref{Fig:5}). At this doping, the $el$-Fermi pockets sink below the Fermi level; however, they are still close enough to contribute to the SC pairing in the $s_\pm$ channel \cite{Boker2017}. Therefore, the BTRS dome appears at $x \gtrsim x_{\rm L}$.  

iii) Above $x \sim 0.8$ the topology of the Fermi surface does not change anymore, however $\gamma_{\rm n}$ keeps increase approaching $x = 1$. This behavior is attributed to the increase of the correlation effects with doping \cite{Hardy2016, Medici2014} and to proximity to another QCP detrimental for $T_{\rm c}$ \cite{Eilers2016, Drechsler2017, Drechsler2018}. At this high doping level, the $el$ pockets move further away from the Fermi energy and cannot anymore contribute to the SC pairing. As a result, the order parameter exhibits another type of the sign change a nodal $s$-wave state \cite{Chubukov2015, Ota2014, Watanabe2014, Cho2018}. Thus, the specific heat in the normal state indicates that the BTRS state appears as a result of essential changes of the electronic structure. Our data are consistent with the scenario proposed in Ref.\ \cite{Boker2017}, where the Lifshitz transition of the vanishing $el$ Fermi pockets at $x_{\rm L} \simeq 0.55 - 0.7$ drives the BTRS state at a doping level $x$ slightly larger than $x_{\rm L}$.

In summary, we were able to identify a dome of a novel SC phase in the Ba$_{\rm 1-x}$K$_{\rm x}$Fe$_2$As$_2$ system that breaks time reversal symmetry,  combining $\mu$SR and thermodynamic studies. The analysis of the $\mu$SR data indicates that this phase has an $s+is'$ symmetry of the SC order parameter. The BTRS-dome appears in a narrow doping range of about $x \approx$ 0.7 - 0.85 as an intermediate SC phase between a nodeless and a nodal $s_{\pm}$ SC states. The subtle balance between these pairing states is tuned by the topological changes of the Fermi surface as a function of hole doping. The existence of the BTRS-dome provides an important support for the validity of the theoretical understanding of unconventional multiband superconductivity in iron pnictides. In addition, our work opens a new avenue for future experimental studies of the unusual physical properties of this novel superconducting state. In particular, an analogous BTRS-dome can appear also in other related isovalent compounds A$_{1-x}$M$_x$Fe$_2$As$_2$ close to topological changes of the Fermi surface, with A = Sr, Ca and M = Na, Rb, and  Cs. We expect that the size of the dome may vary significantly between these compounds since the correlation effect and the electronic structure are affected by the size of the cations. Any changes, if detected experimentally, would provide important detailed insights for a still missing general microscopic as well as phenomenological theory of the multiband superconductivity in these materials.

\section*{Methods}

\subsection*{Samples}
The samples used in $\mu$SR measurements consist of the $c$-axis oriented stacks of the plate-like Ba$_{\rm 1-x}$K$_{\rm x}$Fe$_2$As$_2$ single crystals selected according to their $T_{\rm c}$ values. Phase purity and crystalline quality of the crystals were examined by X-ray diffraction (XRD) and TEM. The $c$-lattice parameters were calculated from the XRD data 
using the Nelson Riley function.  The K doping level $x$ of the single crystals was calculated using the relation between the $c$-axis lattice parameter and
the K doping obtained in the previous studies \cite{Kihou2016}. 

\subsection*{Specific heat and magnetization measurements}

The $T_{\rm c}$-values of the samples (consisting of a stack of single crystals) used for the $\mu$SR experiments were defined from DC susceptibility measurements using a commercial superconducting quantum interference device (SQUID) magnetometer from Quantum Design. The $T_{\rm c}$-values of the individual crystals given in Fig.\ \ref{Fig:5} were obtained using the entropy conservation method from the specific heat data measured in a Quantum Design physical property measurement system (PPMS). The phonon contribution in the specific heat was determined experimentally for single crystals with $T_{\rm c} \lesssim 10$ K (see Fig.\ S4 in the SM). For crystals with higher $T_{\rm c}$ (i.e. in the doping range $0.8 \lesssim x \lesssim 0.65$) we used the phonon contribution of the low $T_{\rm c}$ samples with an adjustment coefficient following the procedure proposed in Ref.\ \cite{Hardy2016}.  

\subsection*{$\mu$SR experiments}

The most of the $\mu$SR experiments on the stacks of the Ba$_{\rm 1-x}$K$_{\rm x}$Fe$_2$As$_2$
single crystals were performed at the GPS instrument \cite{GPS}  and the sample with  $x = 0.98(2)$ was measured at the LTF instrument of the $\pi$M3 beamline 
the Paul Scherrer Institute (PSI) in Villigen, Switzerland. Fully spin-polarized positive muons with an energy of 4.2 MeV were implanted in the sample (parallel to the crystallographic $c$-axis). The sample size and sample mounting were similar as in our previous measurements \cite{Grinenko2017}. To optimize the amount of muons stopped in the sample we used an Ag degrader with a thickness of $d_{\rm Ag}$ = 50 $\mu$m. ZF and transverse-field (TF) measurements were performed in the transverse polarization mode, where the muon spin polarization $P_{\rm \mu}$ is at 45$^o$ with respect to the muon beam (pointing toward the upward counter) and the sample $c$-axis \cite{GPS}. Using the veto detectors of the GPS instrument we were able to measure small samples with a mass of 10 - 20 mg. For the measurements at the LTF instrument, we used a bigger  sample with the mass of about 60 mg since this instrument was not equipped with a veto detector system. The depolarization rates of the muon polarization components $P_{\rm \mu} \parallel c$ and $P_{\rm \mu} \parallel a$ were determined from forward-backward and up-down positron detector pairs, respectively. The background contribution to the sample signal was carefully determined for each sample using TF measurements above and below $T_{\rm c}$ as shown in the in the SM. The obtained background contribution were fixed in the analysis of the ZF data. The data were analyzed using the MUSRFIT software package \cite{Suter2012}. Further details of our data analysis can be found in the SM.

\subsection*{Acknowledgment} This work was supported by DFG (GR 4667, GRK 1621, and SFB 1143 project C02). S.-L.D an D.E. thank the VW fundation for financial support. This work was performed partially at the Swiss Muon Source (S$\mu$S), PSI, Villigen. We acknowledge fruitful discussion with A. Amato, S. Borisenko, E. Babaev, A. Charnukha, O. Dolgov, C. Hicks, C. Meingast, and A. de Visser. We are grateful for the technical assistance to C. Baines.

\subsection*{Authors contribution} V.G. designed the study, initiated and supervised the project and wrote the paper, performed the $\mu$SR, specific heat, magnetic susceptibility, x-ray diffraction measurements, R.S. performed the $\mu$SR experiments, K.K. and C.H.L. prepared Ba$_{\rm 1-x}$K$_{\rm x}$Fe$_2$As$_2$ single crystals in the doping range $0.65 \lesssim x \lesssim 0.85$, I.M, S.A.  prepared Ba$_{\rm 1-x}$K$_{\rm x}$Fe$_2$As$_2$ single crystals with $x \sim 0.98$, P.C. performed and W.S. supervised TEM measurements, K.Nenkov performed specific heat and magnetization measurements, B.B, R.H. and K.Nielsch supervised the research at IFW, D.E. and S.L.D. provided the interpretation of the experimental data, V.L.V. and M.A.S. performed calculation of the spontaneous magnetic fields in the BTRS state, P.V. and I.E. analysed specific heat in the BTRS state, H.L. performed $\mu$SR experiments and supervised the research at PSI, H.-H.K. initiated the project and supervised the research. All authors discussed the results and implications and commented on the manuscript.

\subsection*{Competing interests}
The authors declare no competing financial interests.

\subsection*{Data availability} The data that support the findings of this study are available from the corresponding author upon reasonable request.

\subsection*{Additional information}
Supplementary information is available for this paper.

Correspondence and requests should be addressed to V.G.

\newpage
\section*{Supplementary material}

\renewcommand{\theequation}{S\arabic{equation}}
\renewcommand{\thefigure}{S\arabic{figure}}
\renewcommand{\thetable}{S\arabic{table}}
\setcounter{equation}{0}
\setcounter{figure}{0}
\setcounter{table}{0}

{\bf In the Supplementary material we show transmission electron microscopy (TEM), additional $\mu$SR, specific heat, and electrical resistivity data. We also provide details of the calculations of the spontaneous magnetic fields in the broken time reversal symmetry (BTRS) state within a phenomenological model.}

\begin{figure}[b]
\centering
\includegraphics[width=10cm]{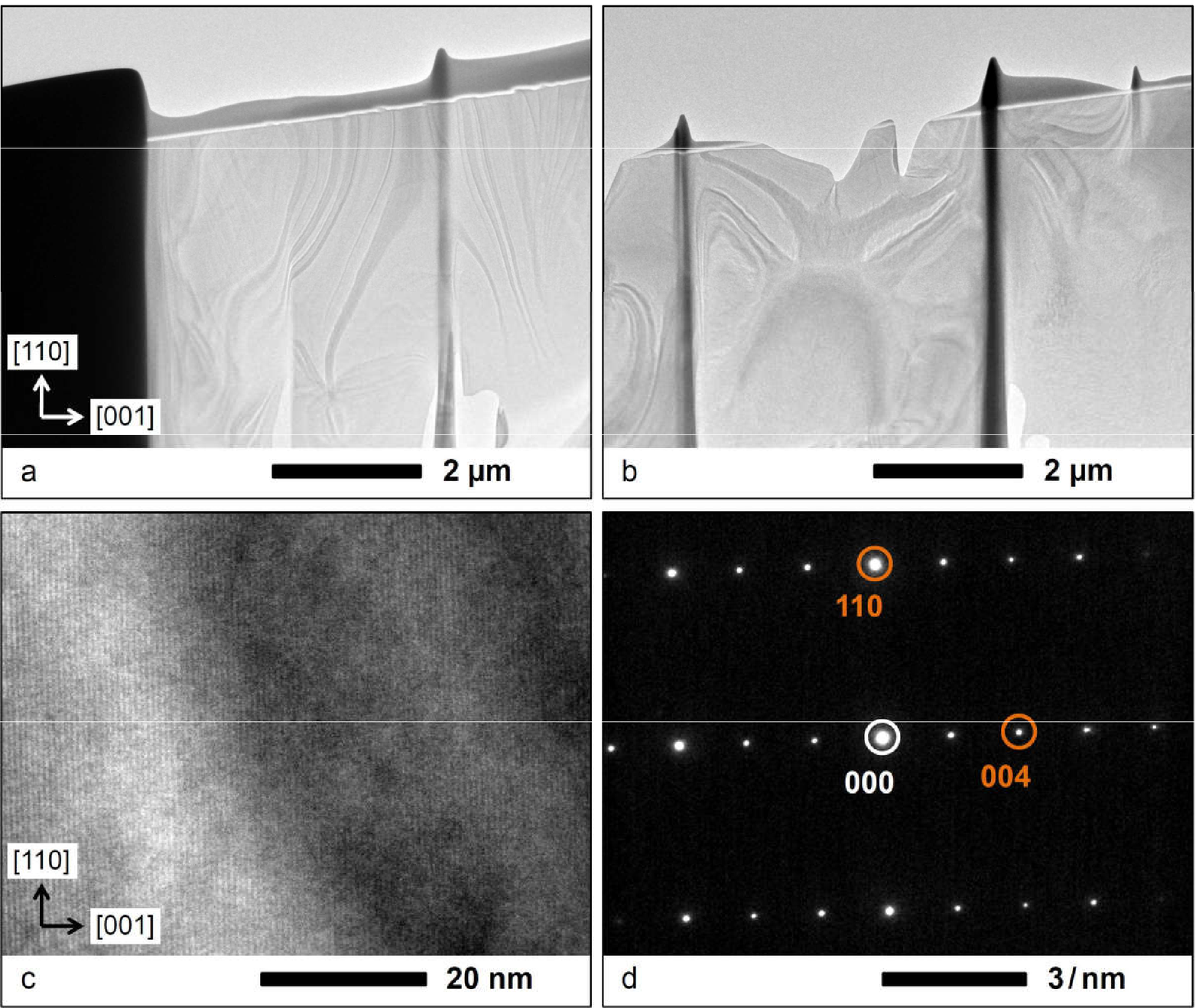}
\caption{TEM overview of Ba$_{\rm 1-x}$K$_{\rm x}$Fe$_2$As$_2$ single crystal with $x =$ 0.74(3) 
according to energy-dispersive X-ray spectroscopy. a) to c) present TEM bright field micrographs of the lamella at different magnifications and d) shows a selected area diffraction pattern.} 
\label{FigS1}
\end{figure} 

\section{TEM}
The TEM lamella was prepared applying the focused ion beam (FIB) technique in a FEI Helios 600i FIB using single crystalline  Ba$_{\rm 1-x}$K$_{\rm x}$Fe$_2$As$_2$ after the $\mu$SR experiments. The cut was done in  a way that the electron beam is parallel to a $<110>$ direction of the sample. Figures\ \ref{FigS1} a) to c) show typical TEM bright field micrographs of the sample. Bending and thickness fringes are visible. Besides the defects caused by the FIB preparation all recorded TEM micrographs do not reveal any impurity phase or other coarse crystalline defects. This is in agreement with the recorded selected area diffraction patterns (e.g. in Fig.\ \ref{FigS1}d), which only contain spots that correspond to the Ba$_{\rm 1-x}$K$_{\rm x}$Fe$_2$As$_2$ phase.

\section{$\mu$SR}

\begin{figure}[b]
\includegraphics[width=25pc,clip]{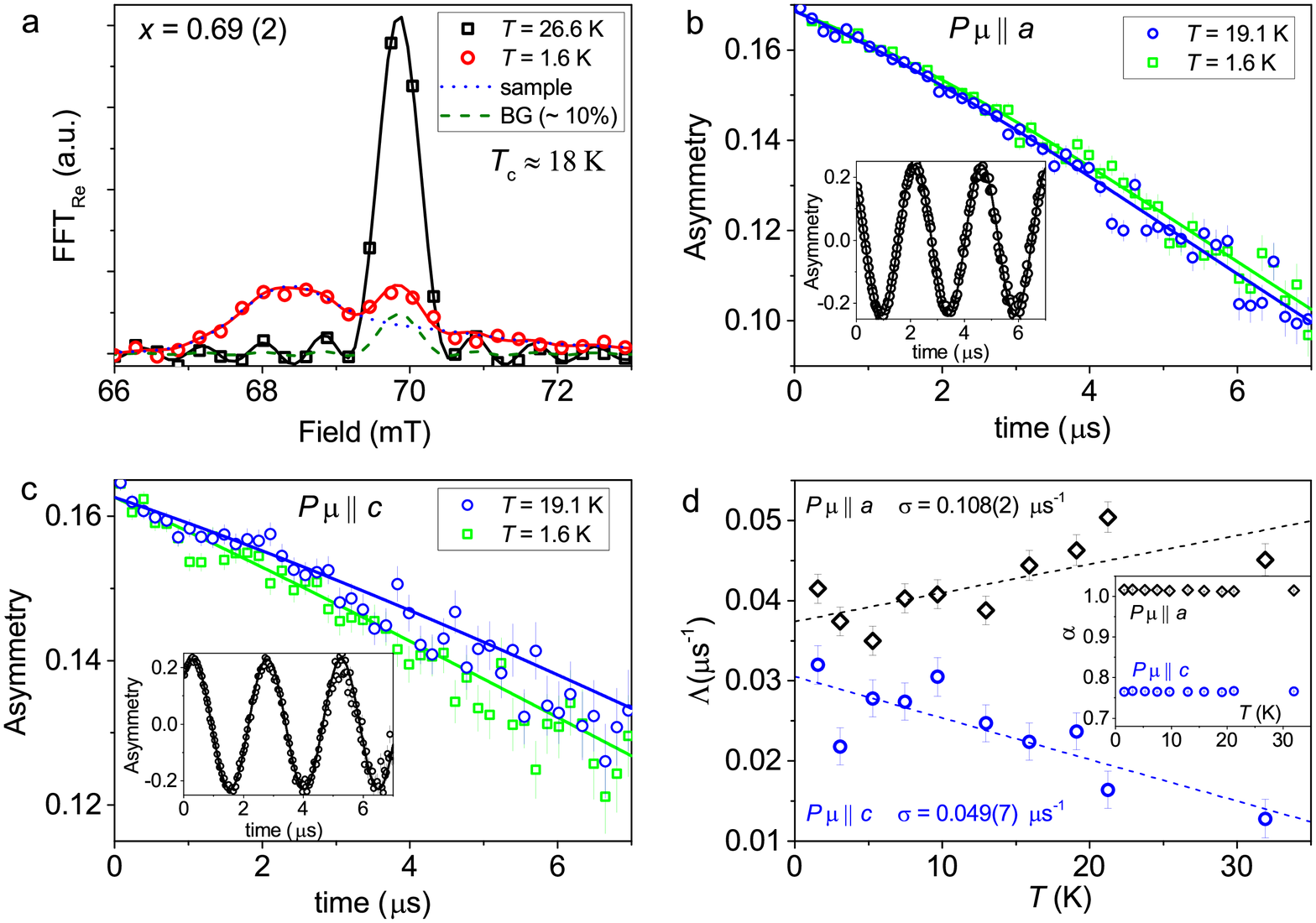}
\caption{The summary of the analysis of the $\mu$SR data for Ba$_{1-x}$K$_x$Fe$_2$As$_2$ sample with $x = 0.69(2)$, $m_{\rm s} \approx$ 26 mg a) The Fourier transform of the TF time spectra measured above and below $T_{\rm c}$. The data in the SC state were measured after cooling in the applied magnetic field. The unshifted part of the spectra is a background (BG) contribution used in Eq.\ \ref{Eq1}. b), and c) Examples of the  ZF-$\mu$SR time spectra at different temperatures for the muon spin component $P_{\rm \mu} \parallel a$ and $P_{\rm \mu} \parallel c$, correspondingly. Symbols - experimental data, solid lines - fits using Eq. \ref{Eq1} ({\bf Strategy 1}). The initial asymmetry $A_{\rm s}(0)$ is obtained from the low TF measurements shown in the insets (according to {\bf Strategy 1} in the SM text). d) The resulting temperature dependencies of the muon spin depolarization rate $\Lambda$ for the two muon spin components obtained within Strategy 1. The dashed curves are the linear fits. Inset: Temperature dependencies of the $\alpha$ values for the two muon spin components.} 
\label{FigS_muSR_1_18K}
\end{figure} 

\newpage

\begin{figure}[t]
\includegraphics[width=25pc,clip]{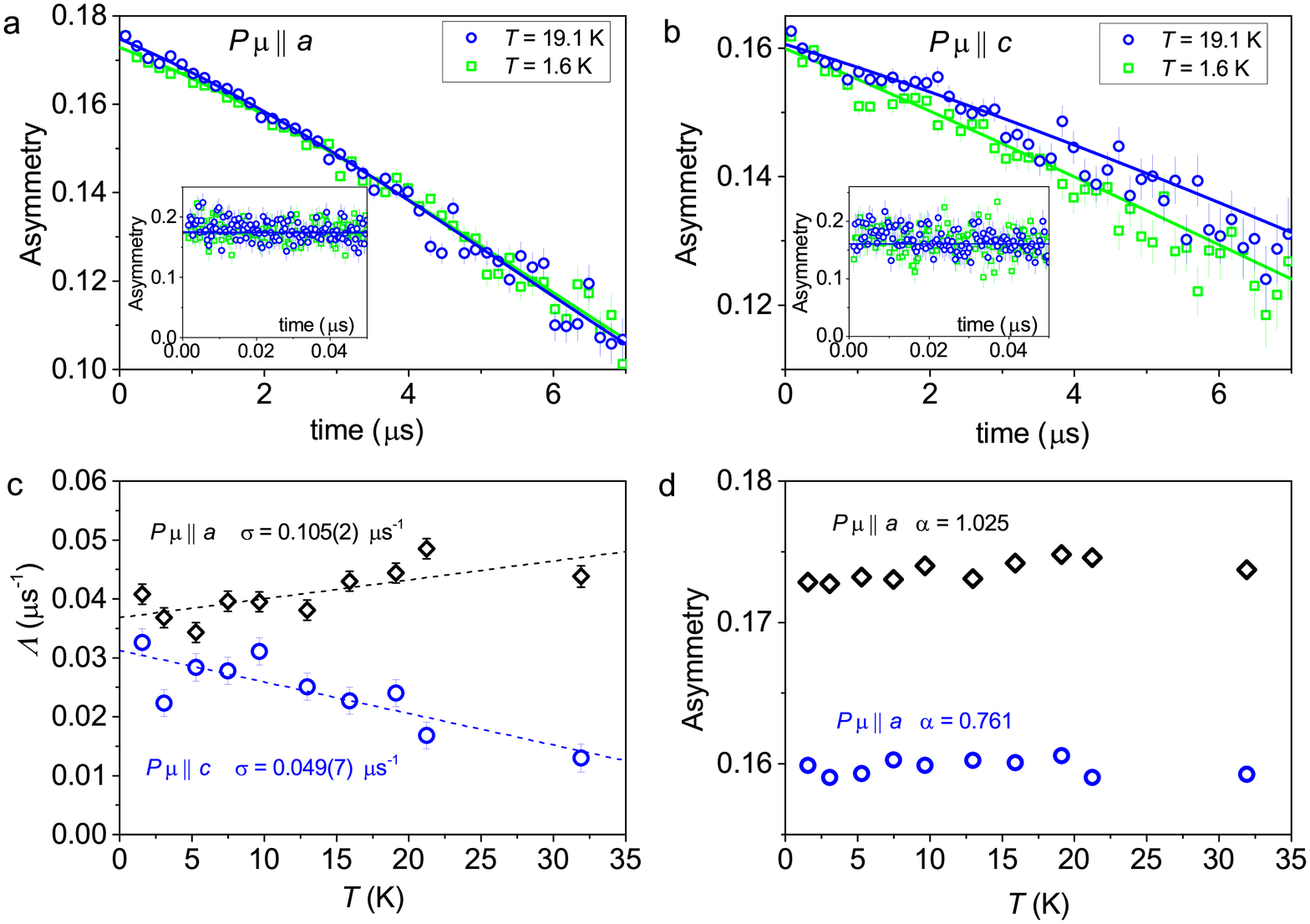}
\caption{The summary of the analysis (using {\bf Strategy 2}) of the $\mu$SR data for Ba$_{1-x}$K$_x$Fe$_2$As$_2$ sample with $x = 0.69(2)$, $m_{\rm s} \approx$ 26 mg. a), and b) Examples of the  ZF-$\mu$SR time spectra at different temperatures for the muon spin component $P_{\rm \mu} \parallel a$ and $P_{\rm \mu} \parallel c$, correspondingly. Symbols - experimental data, solid lines - fits using Eq. \ref{Eq1}. The initial asymmetry $A_{\rm s}(0)$ is obtained  by fixing the $\alpha$ defined using the low TF measurements shown in the insets of Figs.\ \ref{FigS_muSR_1_18K}b, and \ref{FigS_muSR_1_18K}b. Insets: Zoom into the first 0.05 $\mu$s. c) The resulting temperature dependencies of the muon spin depolarization rate $\Lambda$ for the two muon spin components, and d) temperature dependencies of the $A_{\rm s}(0)$ values for the two muon spin components. The dashed curves are are linear fits.} 
\label{FigS_muSR_2_18K}
\end{figure}

\begin{figure}[t]
\includegraphics[width=25pc,clip]{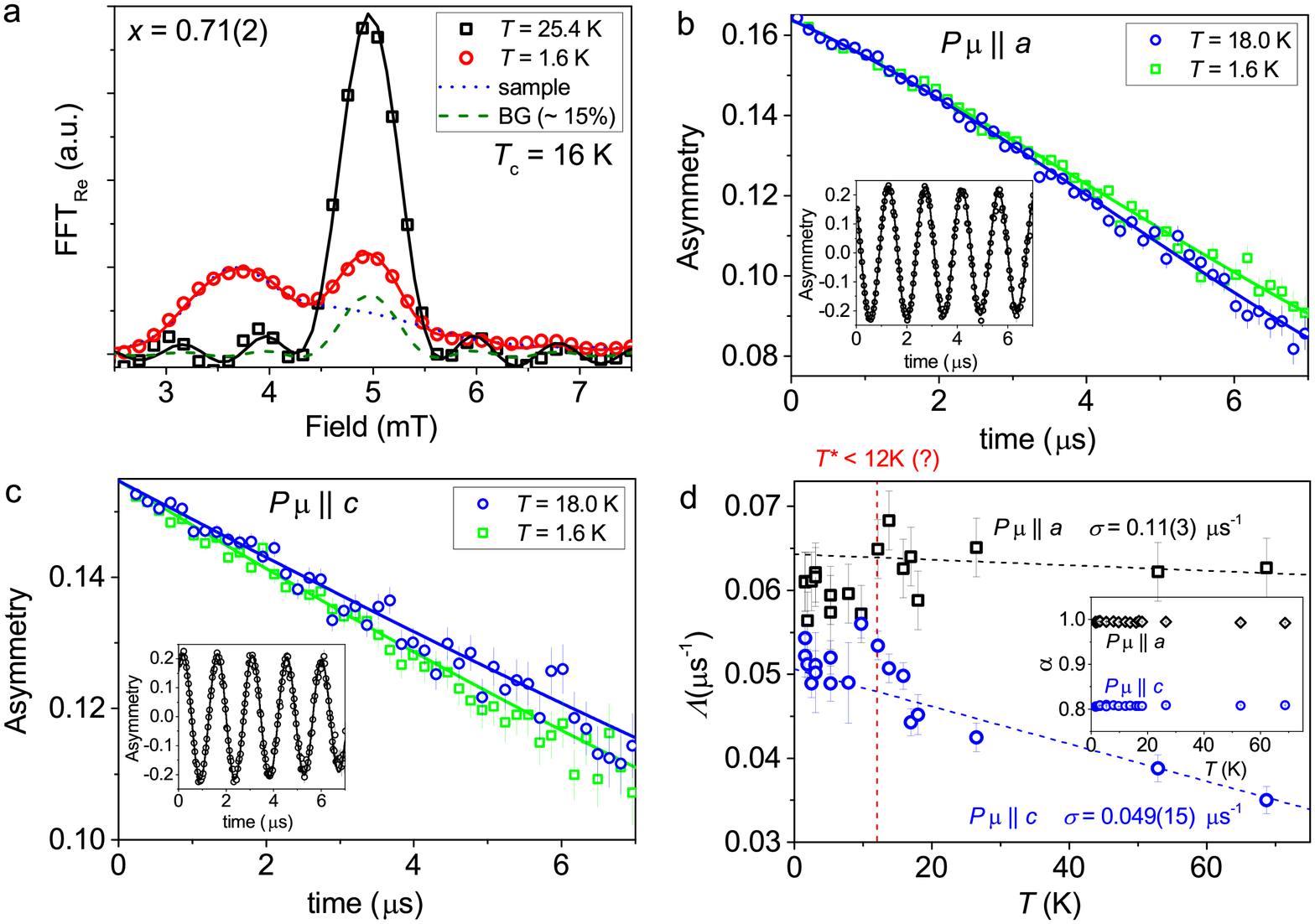}
\caption{The summary of the analysis of the $\mu$SR data for Ba$_{1-x}$K$_x$Fe$_2$As$_2$ sample with $x = 0.71(2)$, $m_{\rm s} \approx$ 20 mg. For the description of the figure sections see Fig.\ \ref{FigS_muSR_1_18K}} 
\label{FigS_muSR_1_16K}
\end{figure} 

\begin{figure}[b]
\includegraphics[width=25pc,clip]{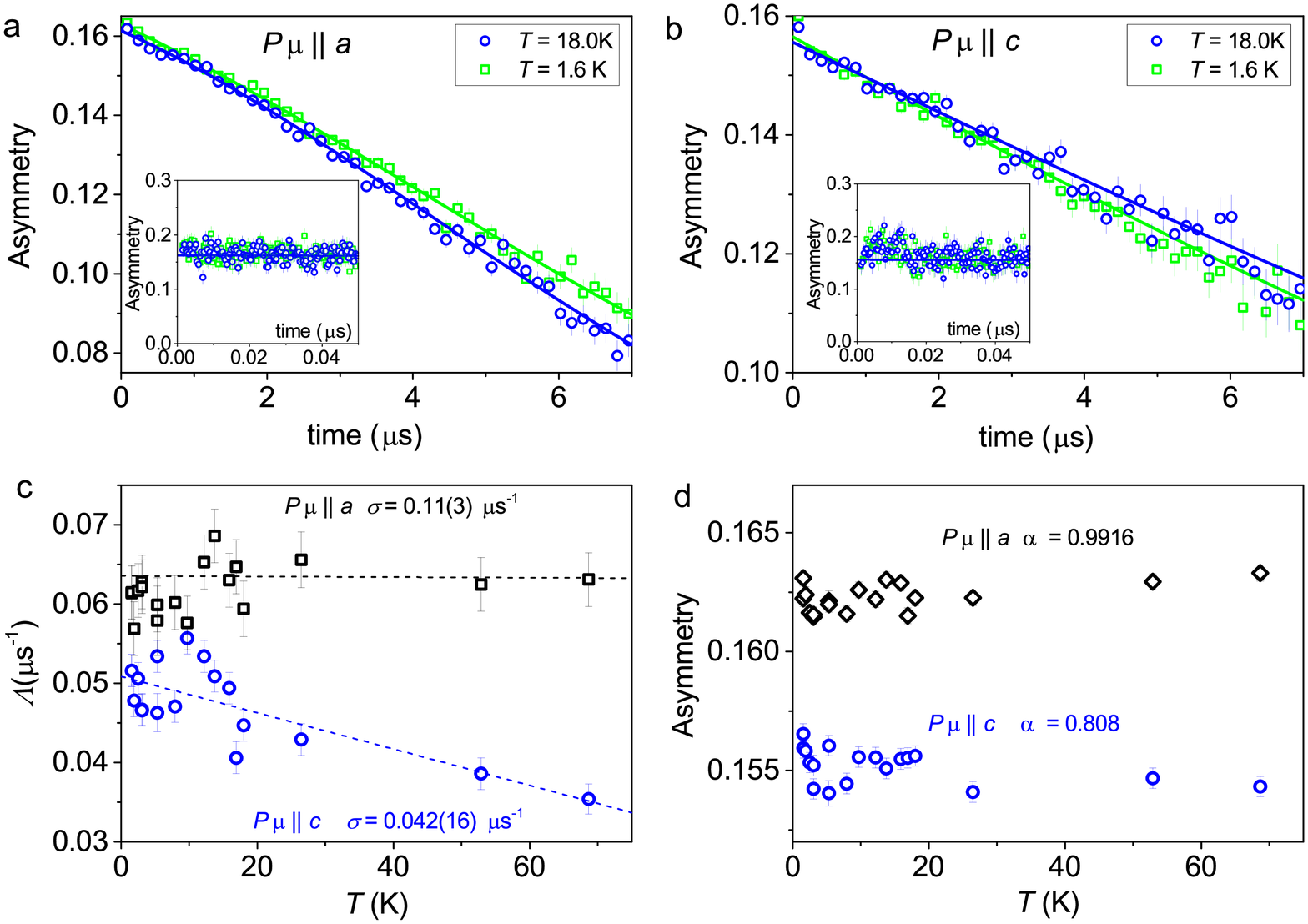}
\caption{The summary of the analysis (using {\bf Strategy 2}) of the $\mu$SR data for Ba$_{1-x}$K$_x$Fe$_2$As$_2$ sample with $x = 0.71(2)$, $m_{\rm s} \approx$ 20 mg. For the description of the figure sections see Fig.\ \ref{FigS_muSR_2_18K}.} 
\label{FigS_muSR_2_16K}
\end{figure}

\begin{figure}[t]
\includegraphics[width=25pc,clip]{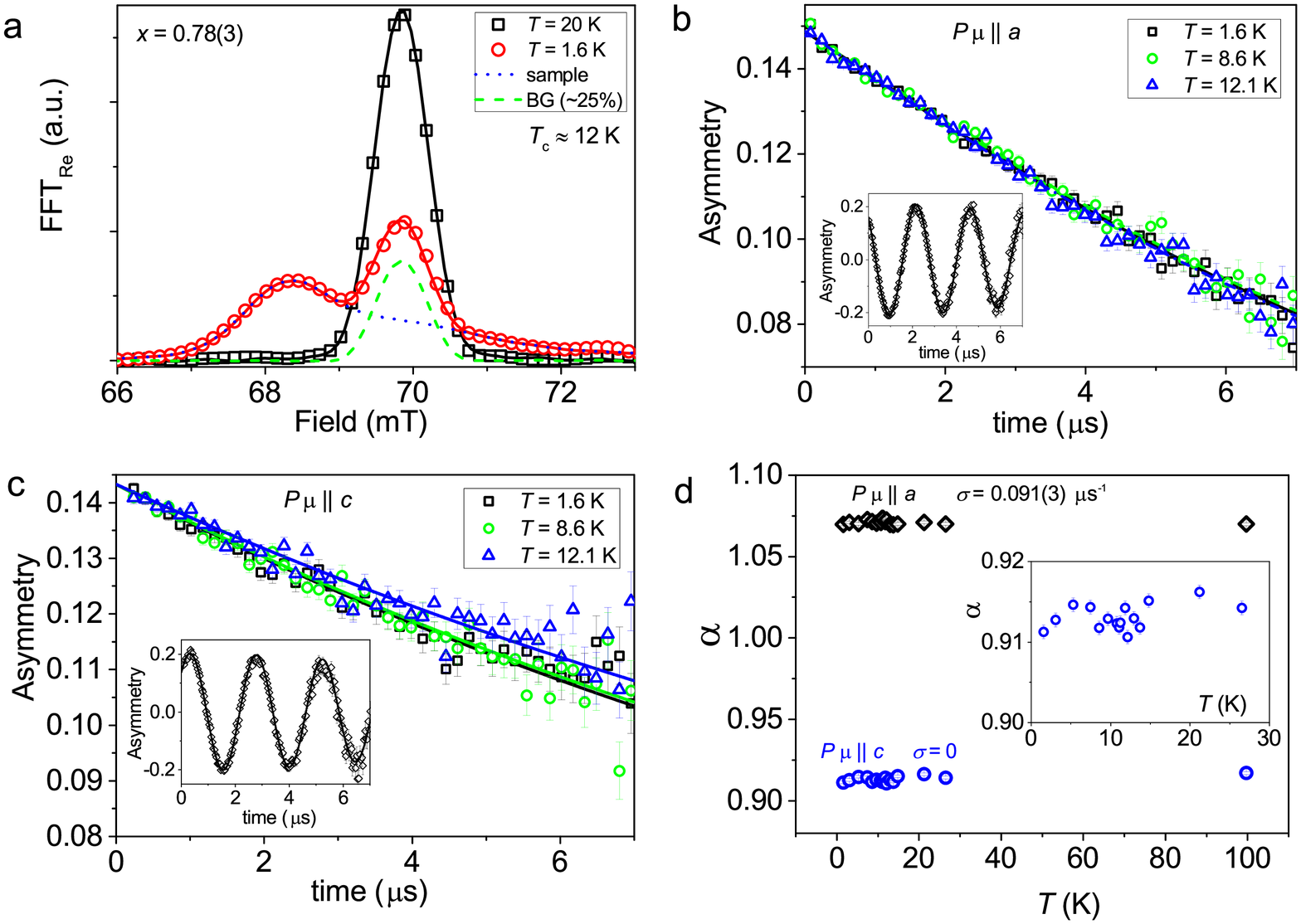}
\caption{The summary of the analysis of the $\mu$SR data for Ba$_{1-x}$K$_x$Fe$_2$As$_2$ sample with $x = 0.78(3)$, $m_{\rm s} \approx$ 22 mg. For the description of the figure sections a)-c) see Fig.\ \ref{FigS_muSR_1_18K}. d) Temperature dependencies of the $\alpha$ values for the muon spin component $P_{\rm \mu} \parallel a$ and $P_{\rm \mu} \parallel c$ obtained using {\bf Strategy 1}. Inset: Zoom into the low-temperature range. } 
\label{FigS_muSR_1_12K}
\end{figure} 

\begin{figure}[b]
\includegraphics[width=25pc,clip]{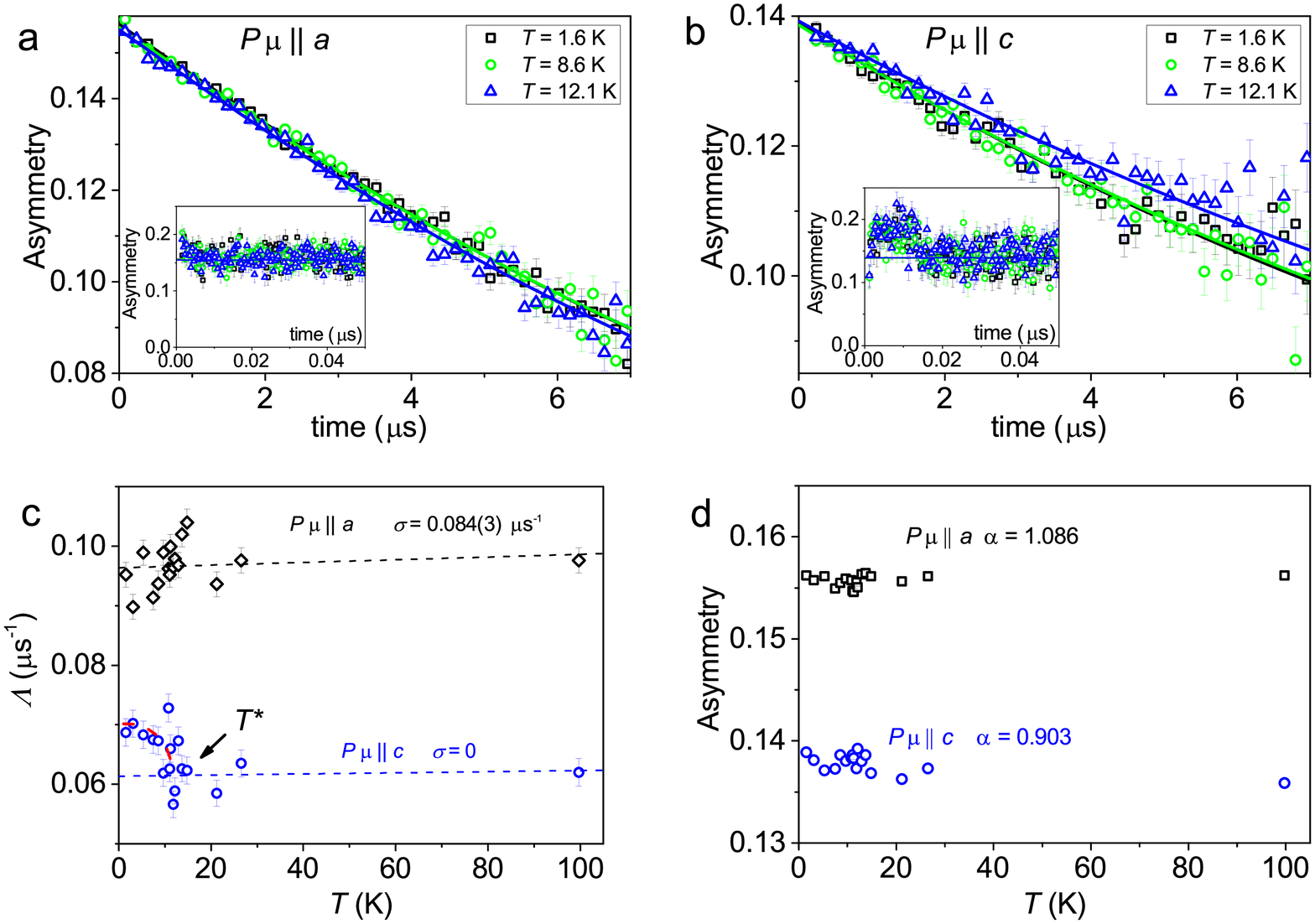}
\caption{The summary of the analysis (using {\bf Strategy 2}) of the $\mu$SR data for Ba$_{1-x}$K$_x$Fe$_2$As$_2$ sample with $x = 0.78(3)$, $m_{\rm s} \approx$ 22 mg. For the description of the figure sections see Fig.\ \ref{FigS_muSR_2_18K}.} 
\label{FigS_muSR_2_12K}
\end{figure} 

\begin{figure}[t]
\includegraphics[width=25pc,clip]{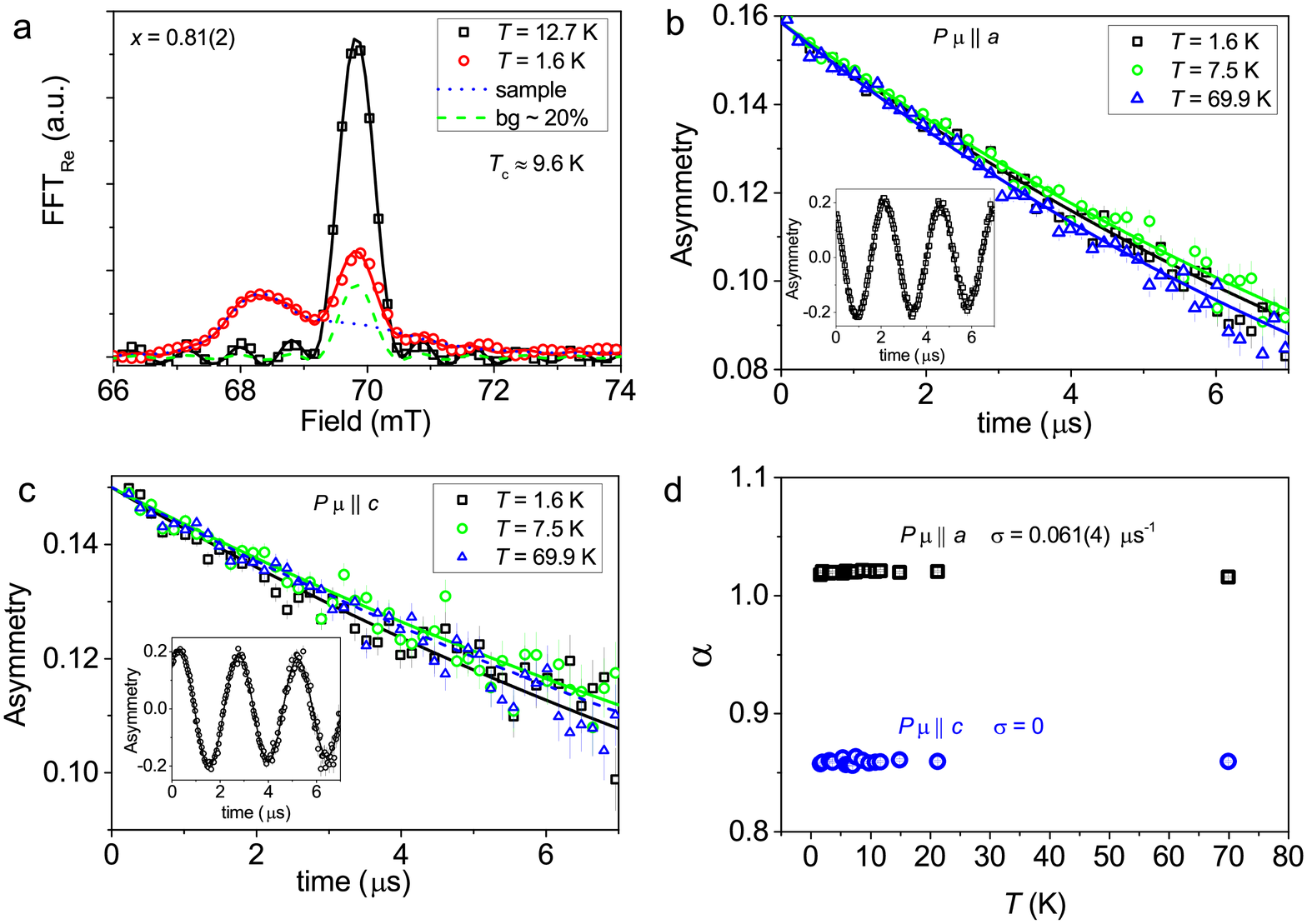}
\caption{The summary of the analysis of the $\mu$SR data for Ba$_{1-x}$K$_x$Fe$_2$As$_2$ sample with $x = 0.81(2)$, $m_{\rm s} \approx$ 23 mg. For the description of the figure sections a)-c) see Fig.\ \ref{FigS_muSR_1_18K}. d) Temperature dependencies of the $\alpha$ values for the muon spin component $P_{\rm \mu} \parallel a$ and $P_{\rm \mu} \parallel c$ obtained using {\bf Strategy 1}.} 
\label{FigS_muSR_1_10K}
\end{figure} 

\begin{figure}[b]
\includegraphics[width=25pc,clip]{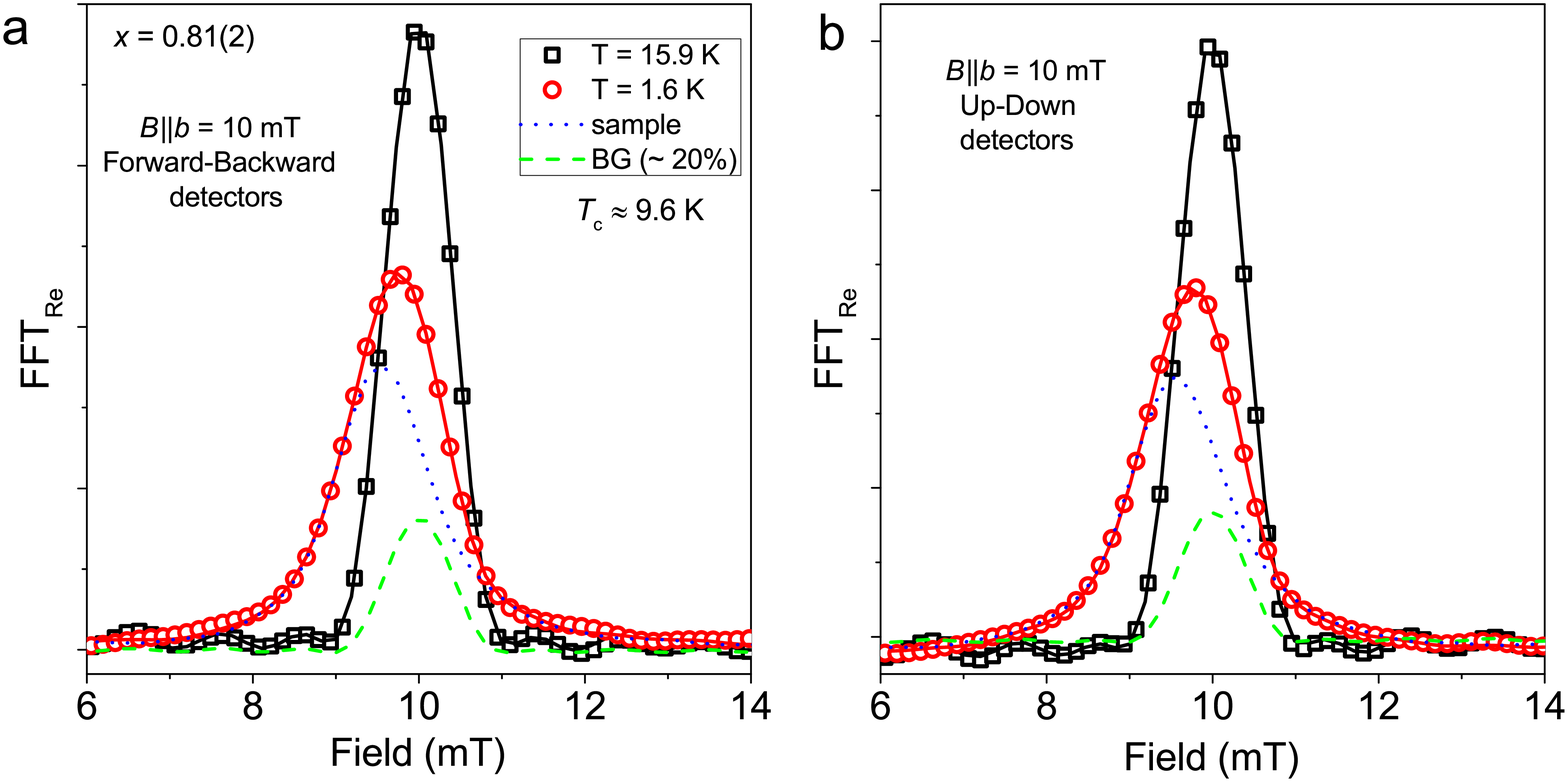}
\caption{The Fourier transform of the TF time spectra measured above and below $T_{\rm c}$ for two orthogonal detector pairs used for collecting ZF data (Ba$_{1-x}$K$_x$Fe$_2$As$_2$ sample with $x = 0.81(2)$, $m_{\rm s} \approx$ 23 mg). The data in the SC state were measured after cooling in the applied magnetic field. The background (BG) contribution is the same for Forward-Backward and Up-Down detector pairs.} 
\label{FigS_10K_SC_raction_2_der}
\end{figure}

\begin{figure}[t]
\includegraphics[width=25pc,clip]{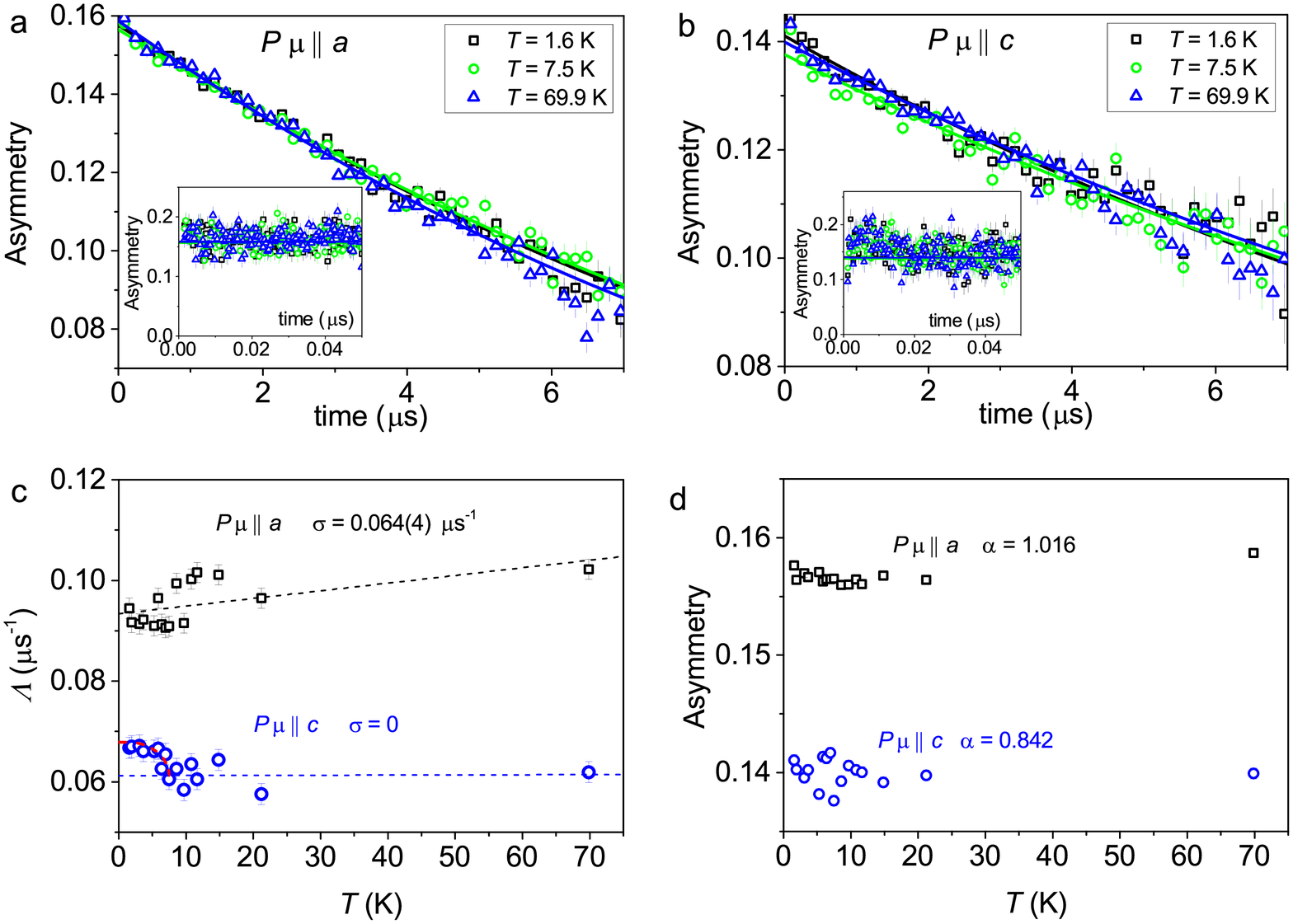}
\caption{The summary of the analysis (using {\bf Strategy 2}) of the $\mu$SR data for Ba$_{1-x}$K$_x$Fe$_2$As$_2$ sample with $x = 0.81(2)$, $m_{\rm s} \approx$ 23 mg. For the description of the figure sections see Fig.\ \ref{FigS_muSR_2_18K}.} 
\label{FigS_muSR_2_10K}
\end{figure}

\begin{figure}[b]
\includegraphics[width=25pc,clip]{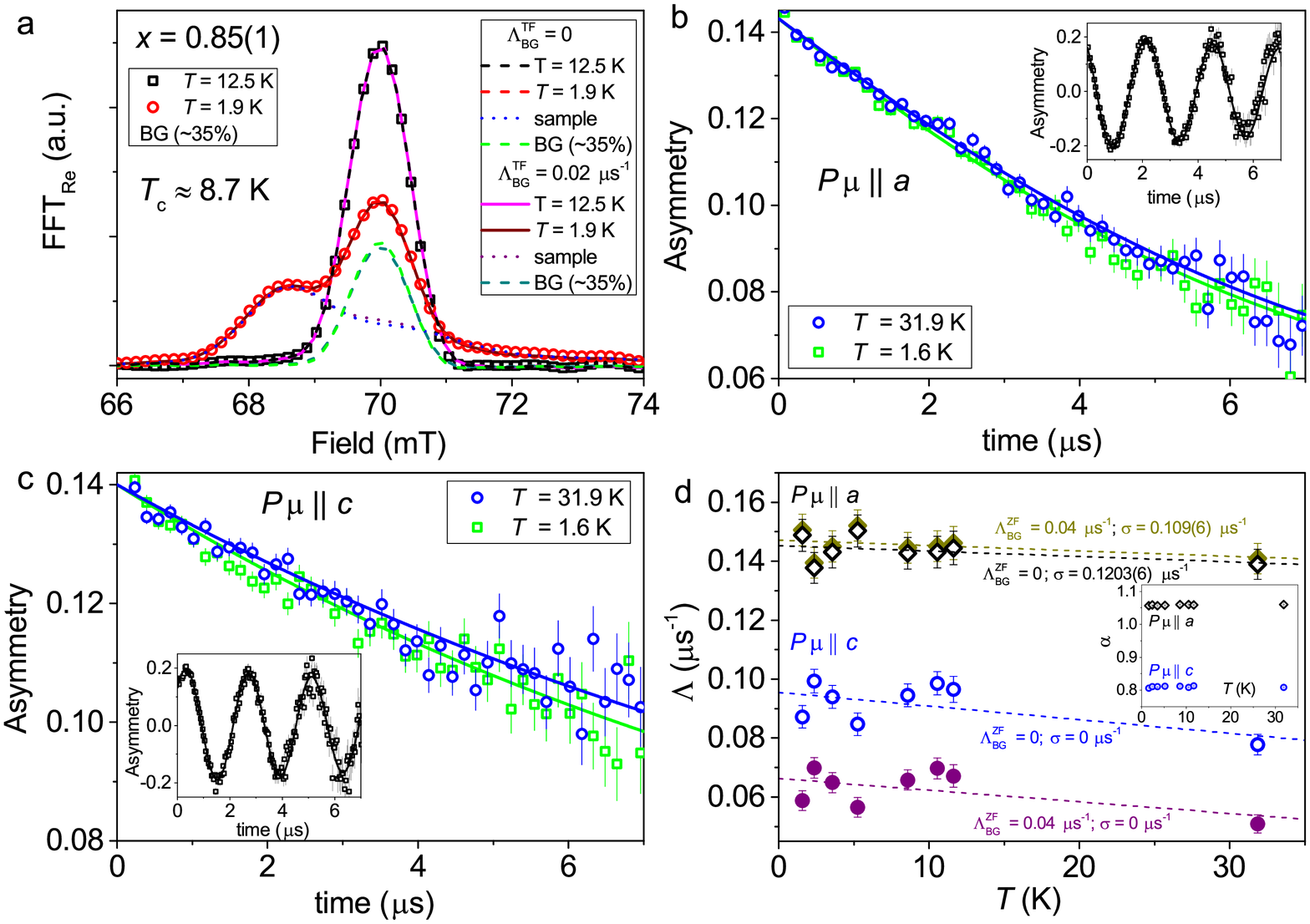}
\caption{The summary of the analysis of the $\mu$SR data for Ba$_{1-x}$K$_x$Fe$_2$As$_2$ sample with $x = 0.85(1)$, $m_{\rm s} \approx$ 9 mg. For the description of the figure sections see Fig.\ \ref{FigS_muSR_1_18K}} 
\label{FigS_muSR_1_9K}
\end{figure} 

\begin{figure}[t]
\includegraphics[width=25pc,clip]{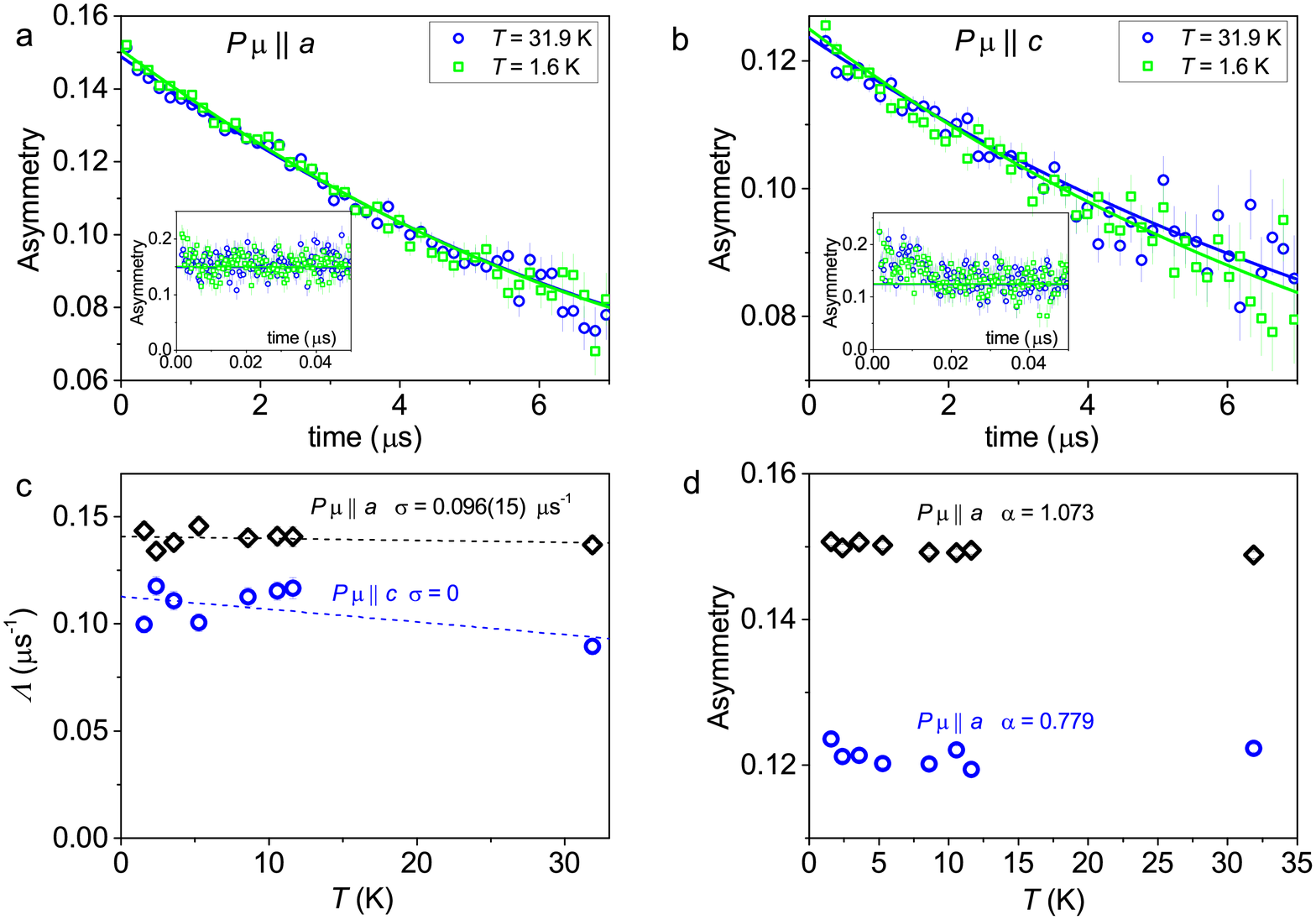}
\caption{The summary of the analysis (using {\bf Strategy 2}) of the $\mu$SR data for Ba$_{1-x}$K$_x$Fe$_2$As$_2$ sample with $x = 0.85(1)$, $m_{\rm s} \approx$ 9 mg. For the description of the figure sections see Fig.\ \ref{FigS_muSR_2_18K}.} 
\label{FigS_muSR_2_9K}
\end{figure} 

\begin{figure}[t]
\includegraphics[width=25pc,clip]{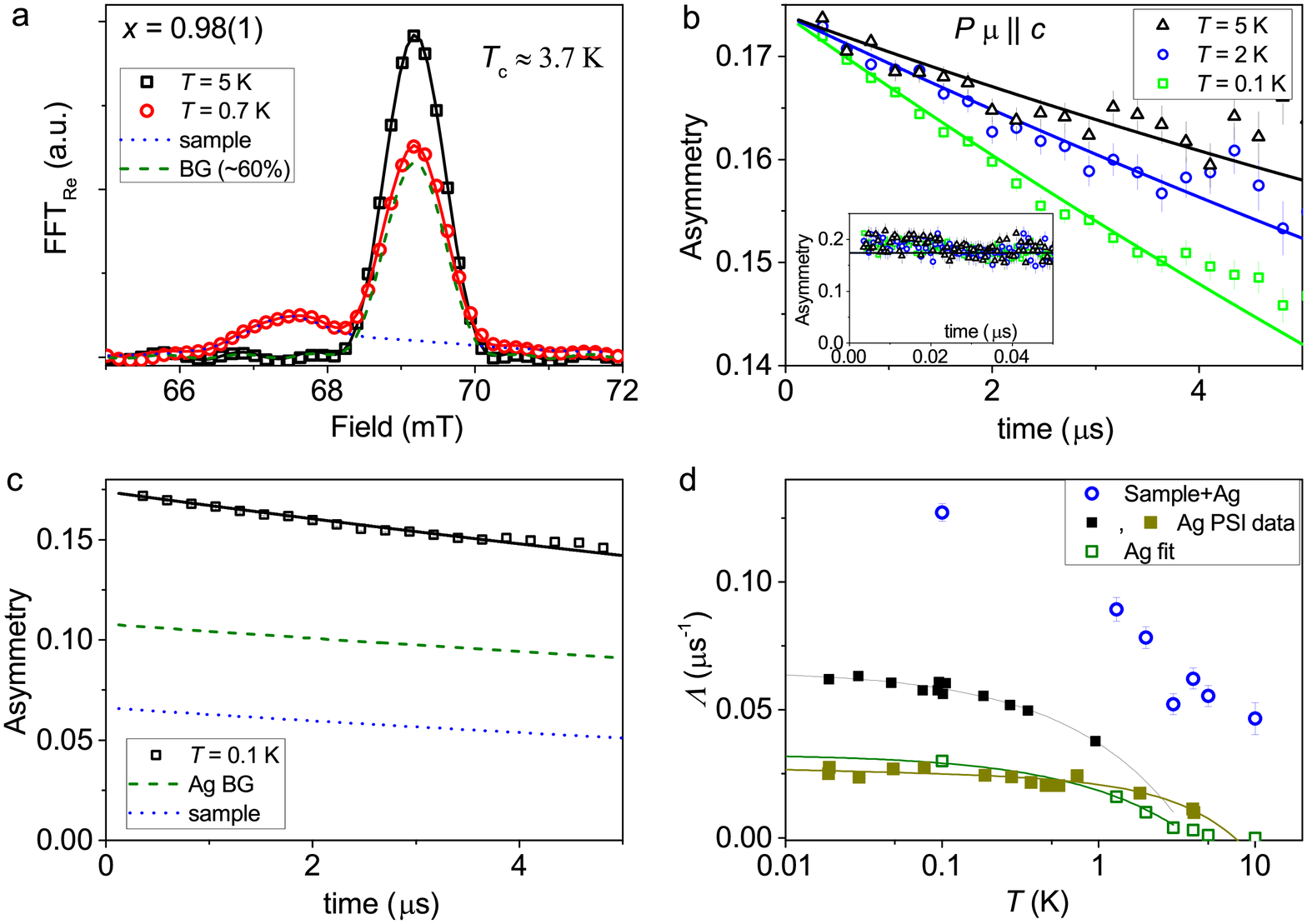}
\caption{The summary of the analysis of the $\mu$SR data for Ba$_{1-x}$K$_x$Fe$_2$As$_2$ sample with $x = 0.98(2)$, $m_{\rm s} \approx$ 60 mg measured in the dilution fridge of the LTF spectrometer. a) The Fourier transform of the TF time spectra measured above and below $T_{\rm c}$. The data in the SC state is measured after cooling in the applied field. The unshifted part of the spectra is a background (BG) contribution in Eq.\ \ref{Eq1}. b) Examples of the  ZF-$\mu$SR time spectra at different temperatures for the muon spin component $P_{\rm \mu} \parallel c$ (analyzed using {\bf Strategy 1}). Symbols - experimental data, solid lines - fits using Eq. \ref{Eq1}. Insets: Zoom to the first 0.05 $\mu$s. c) The ZF-$\mu$SR time spectrum at $T$ = 0.1 K. The dashed and doted curves show contributions related to muons stopping in the Ag holder and sample, correspondingly. d) The temperature dependencies of the muon spin depolarization rate $\Lambda$. Open circles are the total depolarization rate of the sample and Ag holder, squares are examples of the typical depolarization rate of the Ag holders measured in the independent experiments, green open squares are the Ag depolarization rate used to deduce the sample depolarization rate shown in Fig.\ \ref{Fig:3}f in the main text.} 
\label{FigS_muSR_1_4K}
\end{figure}

\begin{figure}[t]
\includegraphics[width=20pc,clip]{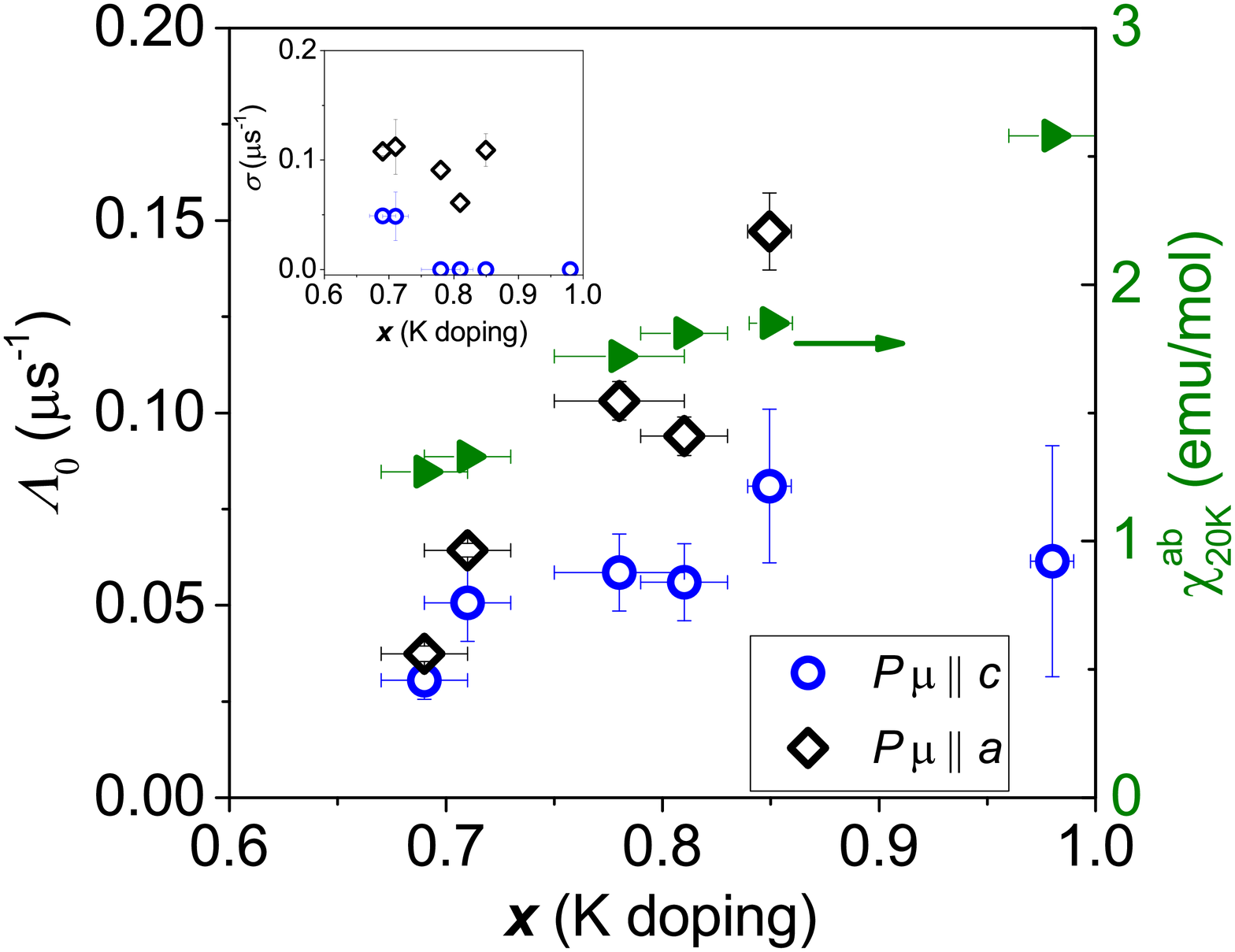}
\caption{(Right axis) Doping dependence of the normal state depolarization rate $\Lambda_0$ extrapolated to zero temperature as shown in Fig. \ref{Fig:3} in the main text obtained using {\bf Strategy 1} described in the SM text. The large error-bars in the absolute value of the depolarization rate for the sample with $x = 0.98(2)$ is caused by uncertainty in the depolarization rate of the Ag contribution (see Fig.\ \ref{FigS_muSR_1_4K}). (Left axis) Doping dependence of the static susceptibility measured at $T = 20$ K. Inset: Doping dependence of the temperature independent depolarization rate $\sigma$ (see Eq.\ref{Eq1}).} 
\label{FigS_ZF_relax_t0}
\end{figure}

The exponential muon spin depolarization rate $\Lambda$ was obtained from ZF-$\mu$SR asymmetry time spectra of the Ba$_{1-x}$K$_{x}$Fe$_2$As$_2$ single crystals using the simplest possible model:
\begin{equation}
A(t) = A_{\rm s}^{\rm ZF}(0){\rm exp}[-\Lambda t]{\rm exp}[-\frac{1}{2}(\sigma t)^2]+A_{\rm bg},\label{Eq1}
\end{equation} 
where $A_{\rm s}^{\rm ZF}(0)$ is the initial sample asymmetry, $A_{\rm bg}$ is the  background asymmetry obtained from transversal field (TF) measurements as shown in Figs.\ \ref{FigS_muSR_1_18K}a, \ref{FigS_muSR_1_16K}a, \ref{FigS_muSR_1_12K}a, \ref{FigS_muSR_1_10K}a, \ref{FigS_10K_SC_raction_2_der}, \ref{FigS_muSR_1_9K}a, and \ref{FigS_muSR_1_4K}a for the samples with different doping levels. $\sigma$ is the temperature independent Gaussian muon spin depolarization rate to account the nuclear contribution given in the inset of Fig.\ \ref{FigS_ZF_relax_t0}. In the analysis for all samples we used $\sigma$ as a global parameter, i.e. the common value of $\sigma$ for all runs. The exception was the sample with $x = 0.71(2)$. The data for this sample were difficult to fit consistently within this strategy. In this case the $\sigma$ values scatter by $\pm 20 \%$ (see Fig.\ \ref{FigS_ZF_relax_t0}). 

For the data analysis we considered two different strategies. Within Strategy 1 we fixed the initial asymmetry $A_{\rm s}^{\rm ZF}(0)$ of our samples according to the asymmetry $A_{\rm s}^{\rm TF}(0)$ defined in TF measurements (see insets to Figs.\ \ref{FigS_muSR_1_18K}, \ref{FigS_muSR_1_16K}, \ref{FigS_muSR_1_12K}, \ref{FigS_muSR_1_10K}, and \ref{FigS_muSR_1_9K}). In the case of the transversal muon polarization $A_{\rm s}^{\rm ZF}(0) = {\rm Cos(45^o)} A_{\rm s}^{\rm TF}(0)$. Further we assumed that the $\alpha$-parameter (characteristic of the detectors efficiency and geometry) could hypothetically be slightly temperature dependent. We think that the assumption that the asymmetry is temperature independent is reasonable since we examine a very weak magnetism. In the raw data, we did not find any indication for a fast reduction of the asymmetry and observed slight distortion at early times $t \lesssim 0.02 \mu$s, only  (see insets in Figs.\ \ref{FigS_muSR_2_18K}, \ref{FigS_muSR_2_16K}, \ref{FigS_muSR_2_12K}, \ref{FigS_muSR_2_10K}, \ref{FigS_muSR_2_9K}, and \ref{FigS_muSR_1_4K}). Therefore, the data at $t \lesssim 0.02 \mu$s were excluded from the analysis. On the other hand, the $\alpha$ value can be temperature dependent due to small change of the sample position with respect to the muon beam and the detectors. The depolarization rate $\Lambda$ obtained within Strategy 1 is shown in the main text. 

To demonstrate that our conclusions are independent on the assumption that the asymmetry is temperature independent we used also Strategy 2 for the data analysis. In this case, we fixed $\alpha$ obtained from the TF measurements and used the asymmetry as a fitting parameter. The results of the analysis using Strategy 2 are shown in Figs.\ \ref{FigS_muSR_2_18K}, \ref{FigS_muSR_2_16K}, \ref{FigS_muSR_2_12K}, \ref{FigS_muSR_2_10K}, and \ref{FigS_muSR_2_9K}). Please note, that we find an anomaly at $T^*$ in the temperature dependence of the depolarization rate $\Lambda$ only, and not in any other fit parameter such as $\alpha$ or asymmetry.     

In all measurements, we observed that some fraction of the signal (BG) was related to muons stopping outside the sample. For the data collected at the GPS spectrometer (measurements down to $T \approx 1.6$ K), we found, as expected, that the BG fraction is the same for both detector pairs (Forward-Backward, and Up-Down) used for ZF measurements (Fig.\ \ref{FigS_10K_SC_raction_2_der}), and  temperature independent within a few percents (error bars of our analysis). We found, also, that the TF depolarization rate of the background is close to zero ($\Lambda^{\rm TF}_{\rm BG}$) and
all TF data can be well fitted assuming a zero depolarization rate for the BG. This result is consistent with the assumption that the BG fraction is caused by an incomplete cancellation of the signal originating from the muons stopped outside the sample (the cryostat walls, for example) by the Veto detectors, and the muons stopped in the Ag degrader with a thickness of up to 50 $\mu$m for the smallest sample. 

To describe the field distribution within the SC sample we use three damped Gaussian components as proposed in Ref.\ \cite{Maisuradze2009}. For the sample with the largest BG we show in Fig.\ \ref{FigS_muSR_1_9K} two examples of the analysis with $\Lambda^{\rm ZF}_{\rm BG} = \Lambda^{\rm TF}_{\rm BG} = 0$ and $\Lambda^{\rm ZF}_{\rm BG} \approx 2\Lambda^{\rm TF}_{\rm BG} = 0.04\mu$s$^{-1}$ (maximum possible muon depolarization rate for the BG) to demonstrate that this small BG depolarization rate does not affect our conclusions. It is seen in Fig.\ \ref{FigS_muSR_1_9K}d that a non-zero  $\Lambda^{\rm ZF}_{\rm BG}$ affects mainly the sample  depolarization rates  $\Lambda$ for $P_{\rm \mu} \parallel c$ and $\sigma $ for $P_{\rm \mu} \parallel a$. This results in a slight increase of the anisotropy of $\Lambda$. The effect is weaker for other samples with a smaller BG fractions and in general, does not change qualitatively the doping and temperature dependencies of the depolarization rates. Therefore, in the main text, we show the data obtained with the zero  depolarization rate of the BG.

We observed a much larger BG fraction at the measurements with the low-temperature spectrometer (LTF). This spectrometer was not equipped with Veto detectors to subtract muons passing through the sample. Therefore, in spite of the relatively large sample ($m_{\rm s} \sim 60$ mg) the BG contribution was about 60\% of the total signal (see Fig.\ \ref{FigS_muSR_1_4K}). For this sample, we used 37 $\mu$m of Ag degrader and the sample was placed on a Ag holder. Therefore, in this case, the BG is related to muons stopping in Ag. In contrast to measurements with the GPS spectrometer above 1.6 K, in which the BG depolarization rate is temperature independent, at low temperatures the Ag depolarization rate is non-zero and strongly temperature dependent (see Fig.\ \ref{FigS_muSR_1_4K}d). The temperature dependence of the Ag depolarization rate is attributed to muons trapped by defects \cite{Bueno2011}.  Therefore, the observed increase of the total depolarization rate at low temperatures is related to the temperature dependent Ag contribution (see Fig.\ \ref{FigS_muSR_1_4K}). In the main text (Fig.\ \ref{Fig:3}f) we present the $\Lambda$ values obtained after subtraction the Ag contribution. 

In Fig.\ \ref{FigS_ZF_relax_t0} we summarize the doping dependencies of the depolarization rate $\Lambda_0$ extrapolated to zero temperature, and the temperature independent $\sigma$ obtained within Strategy 1. We found that the doping dependence of $\Lambda_0$ is similar to the static spin susceptibility $\chi_{\rm 20}$ (at $T$ = 20 K) measured just above the SC transition. Both, depolarization rate and susceptibility monotonously increase approaching $x  = 1$ and do not show any features across the BTRS dome. This excludes any special impurity effect or proximity to a magnetic state at $x \sim 0.7 - 0.85$. 

\begin{figure}[t]
\includegraphics[width=20pc,clip]{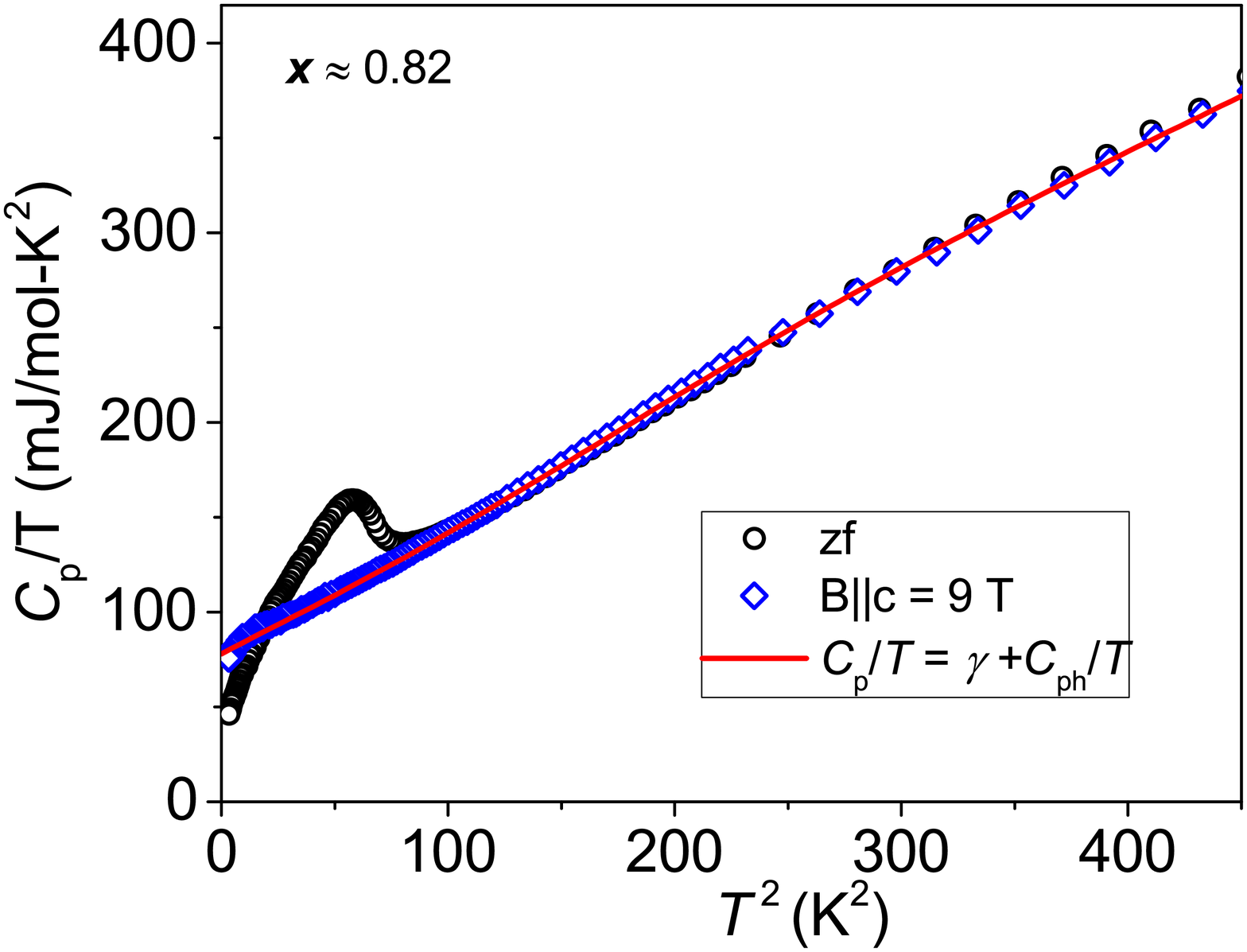}
\caption{Temperature dependence of the specific heat for Ba$_{1-x}$K$_{x}$Fe$_2$As$_2$ single crystal $x \approx 0.82$. Open symbols - measurements in zero and an applied magnetic field of 9 T along the $c$-axis of the crystals. The solid line shows the obtained specific heat in the normal state.} 
\label{FigS4}
\end{figure} 

\section{Specific heat}

The phonon contribution ($C_{\rm ph}$) to the specific heat was determined experimentally using the data in a high magnetic field of the samples with a relatively low $T_{\rm c}\lesssim $10 K as shown in Fig.\ \ref{FigS4} The obtained $C_{\rm ph}$-value multiplied by the coefficient $A = 1 \pm 0.05$ adjusted above $T_{\rm c}$ was used for the samples with a higher $T_{\rm c}$. The quality of the $C_{\rm ph}$ subtraction was examined by the entropy conservation in the superconducting state. This approach is similar to the one used in Ref.\ \cite{Hardy2016s}.

\begin{figure}[t]
\includegraphics[width=20pc,clip]{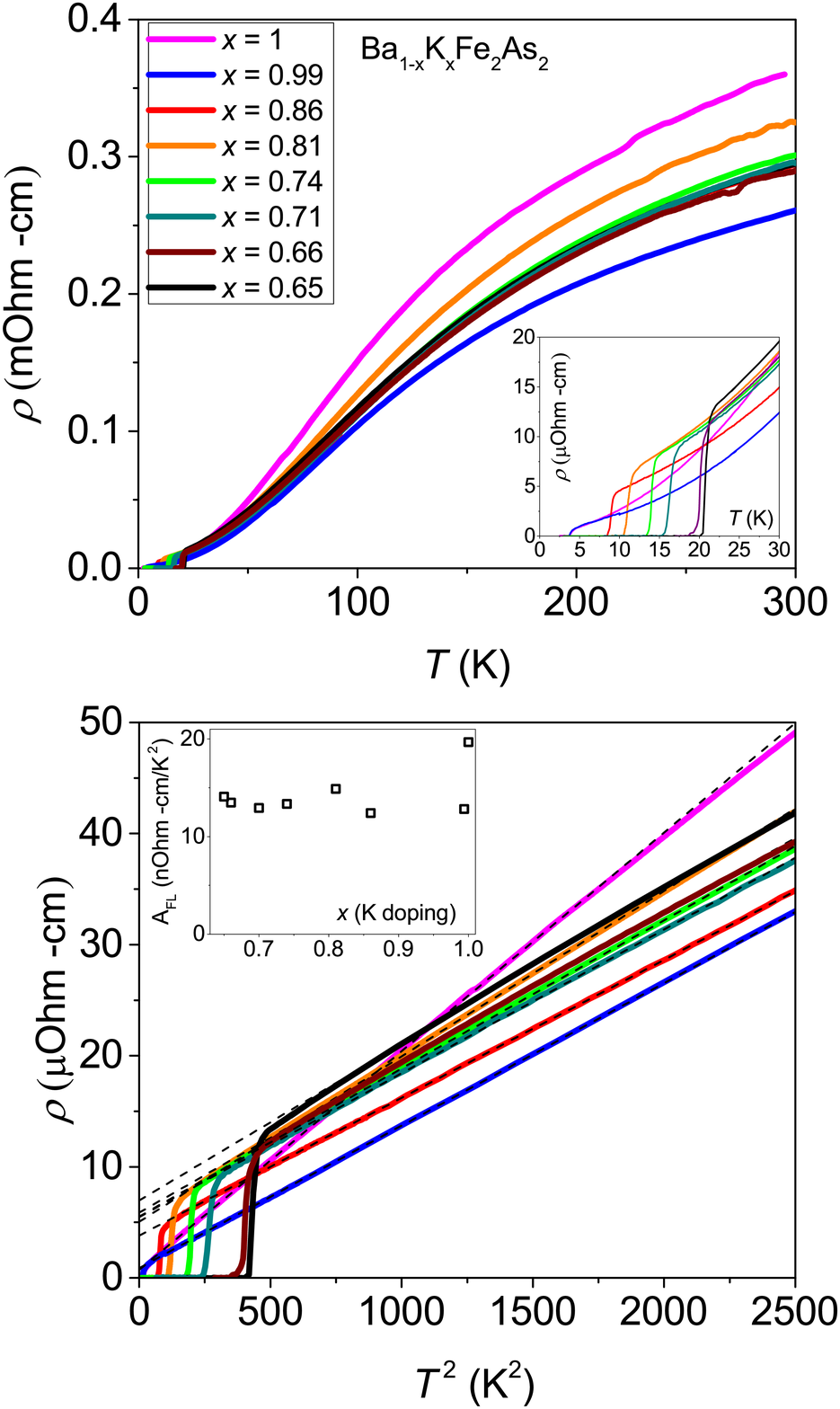}
\caption{ a) Temperature dependence of the resistivity of the Ba$_{1-x}$K$_{x}$Fe$_2$As$_2$ single crystals with different doping levels. Inset: Zoom into the temperature range close $T_{\rm c}$. b) The resistivity versus squared temperature. Dashed curves are the fits using Eq.\ \ref{Eq2}. Inset: Doping dependence of the $A_{\rm FL}$ coefficient. The data for $x = 1$ are taken from Ref. \cite{Grinenko2014a}} 
\label{FigS_res}
\end{figure}

\begin{figure}[t]
\includegraphics[width=26pc,clip]{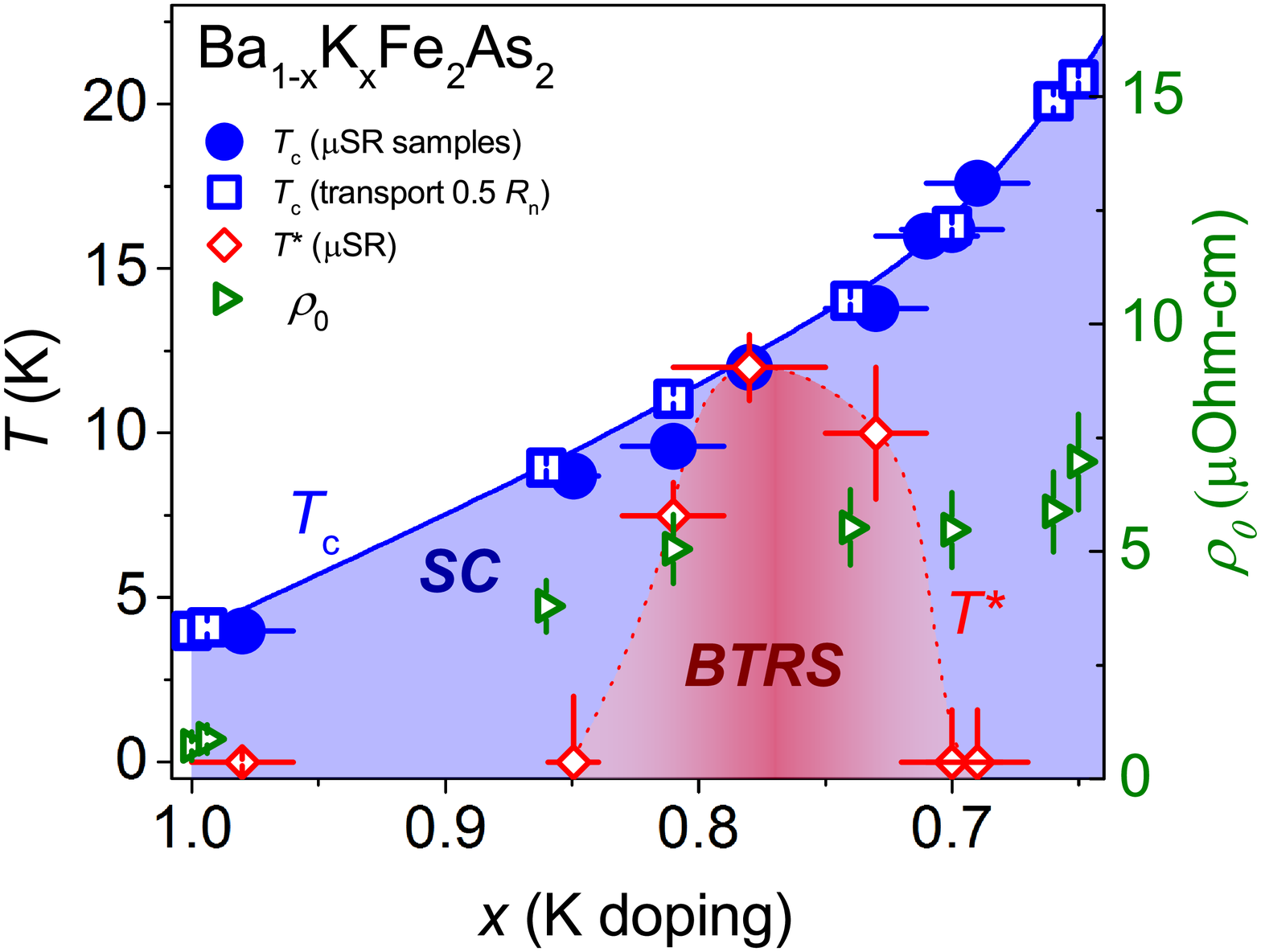}
\caption{ a) Doping dependence of the residual resistivity $\rho_0$ (green triangles, right axis) on top of the phase diagram with the region close to the BTRS dome of the Ba$_{1-x}$K$_{x}$Fe$_2$As$_2$ single crystals. } 
\label{FigS_PD_transport}
\end{figure}

\section{Electrical resistivity}

Temperature dependencies of the electrical resistivity of the Ba$_{1-x}$K$_{x}$Fe$_2$As$_2$ single crystals with different doping levels are summarized in Fig.\ \ref{FigS_res}a. All crystals show a pronounced metallic behavior with a large residual resistivity ratio $RRR = \frac {R(T = 300 {\rm K})} {\rho_0} \gtrsim 50$, where $\rho_0$ is the residual resistivity extrapolated to $T = 0$. At low temperatures the resistivity shows a Fermi liquid (FL) behavior (see Fig.\ \ref{FigS_res}b) and can be described by Eq.\ \ref{Eq2}.

 \begin{equation}
\rho(T) = \rho_0 + A_{\rm FL}T^2,\label{Eq2}
\end{equation}    

We found that the $A_{\rm FL}$ coefficient is nearly doping independent (see in the inset of Fig.\ \ref{FigS_res}b). The doping dependence of $\rho_0$ is shown in Fig.\ \ref{FigS_PD_transport}. The residual resistivity is nearly doping independent in the range $x \sim 0.65 - 0.85$ and then decreases down to $x = 1$. The overall behavior of the transport data, the depolarization rates, and the susceptibility Fig. \ref{FigS_ZF_relax_t0} allow us to conclude that the BTRS state is intrinsic and the increase of the depolarization below $T^*$ is not related to any special kind of disorder, impurity or proximity to a magnetic phase. 

\section{Spontaneous magnetic fields in the BTRS state}

The anisotropy and magnitude of the spontaneous magnetic fields in the BTRS superconducting state 
were estimated using the phenomenological model proposed in Ref.\cite{Vadimov2018s}.

To calculate spontaneous fields in the anisotropic $s+is'$ and $s-is'$ states
generated by the sample inhomogeneities, such as the inhomogeneous doping level distribution 
we have adopted  the following Ginzburg-Landau free energy
derived in \cite{Garaud2017}. The intrinsic sample inhomogeneity parametrized by the  interband 
pairing constant $\eta_2=\eta_2(\bm r)$ shows up in the spatially dependent 
coefficients of the Ginzburg-Landau functional.

 The structure and magnitude of spontaneous magnetic fields generated by such sample inhomogeneities 
  are determined by the anisotropy of the gradient terms coefficients given by 
  $\hat K_{i} = \langle \bm v_{i} \bm v_{i} \rangle$ where 
  $\bm v_{i} $  is the normalized Fermi velocity in the $i$-th superconducting band.  
 For the $s+is'$ state that is isotropic in the $ab$ plane the gradient components 
 are related by the following symmetry relations $K_{i}^x=  K_{i}^y\equiv =  K_{i}^{xy}$. 
 However, for the system being inhomogeneous along the $z$ direction as well as in the 
 $x,y$ directions, we have to take into account the crystal anisotropy which 
 yields in general $K_{i}^z\neq K_{i}^{xy}$.

 This anisotropy is very important due to the following reason.   
 In general, the inhomogeneous pairing coefficients  in $s+is^\prime$ 
 state  produce local phase shifts of the interdand phase differences 
 thus giving rise to the partial currents in each superconducting  band. These currents are proportional to 
 the phase gradients of the   
 corresponding  gap function  $\Delta_{1,2,3}$. 
  Due to the anisotropy partial currents cannot compensate each other and sum up into the non-zero total current.
   We can illustrate the basic properties of
   spontaneous field and current considering the example of 3D spherically symmetric inhomogeneities. 
 Following the work Ref.(\cite{Vadimov2018s}) we use analytical expression obtained in the regime when the scale of inhomogeneity 
  is much larger than the London penetration length. In this case adopting the equations from Ref.(\cite{Vadimov2018s}) we get the magnetic field given by

   \begin{figure}
   \centering
\includegraphics[width=12cm]{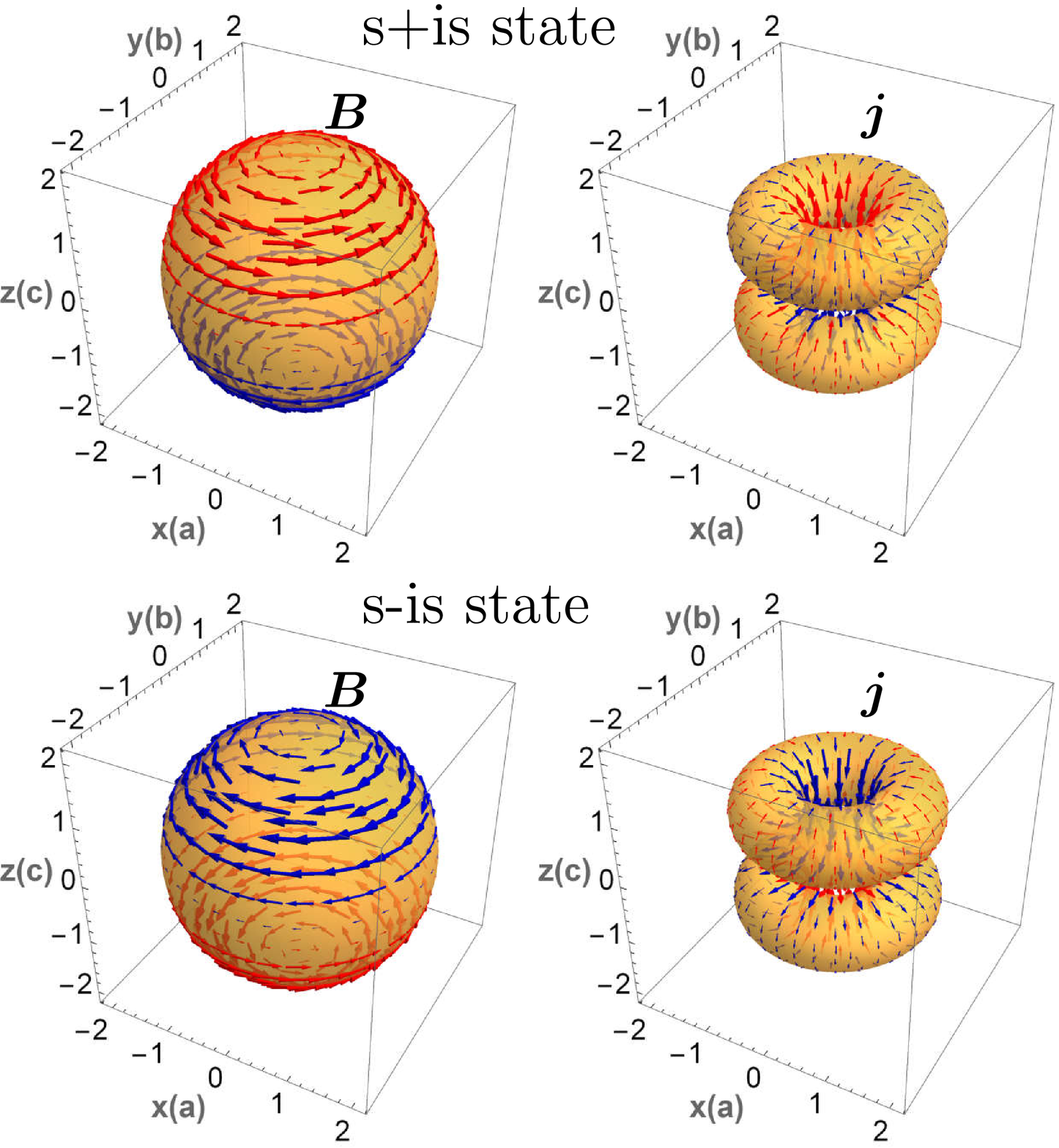}    
\caption{
The structure of the spontaneous magnetic field (left panels) and spontaneous currents (right panels) produced by the 
spherically-symmetric inhomogeneity in the anisotropic $s+is$ and $s-is$ states. 
In the left panels the red/blue arrows show clockwise/counter-clockwise parts of the magnetic field distribution.
The clockwise and counter-clockwise field is generated by the supercurrents with $j_z>0$ (red arrows) and $j_z<0$ 
(blue arrows) shown in the right panels. Notice that the magnetic field and current directions are opposite in the $s+is$ and the $s-is$ states. 
} 
\label{FigS7}
\end{figure}
   
      %
   \begin{align}\label{Eq:mfGL1x}
    &  B_x = B_0
    (k_{12}^{xy}-k_{12}^z) \nabla_{zy} \eta_2 (\bm r)
   \\ \label{Eq:mfGL1y}
    &  B_y = B_0
    (k_{12}^z-k_{12}^{xy}) \nabla_{zx} \eta_2  (\bm r) ,
     \end{align}      
     where $\hat k_{12} = \hat K_1 -\hat K_2$ and $B_0$ is the amplitude which depends of the parameters of the model. 
          The component $B_z$ is zero within the linear approximation that we made, that is 
     when only the phase difference changes in space without affecting the gaps amplitudes. 
    The current which creates this magnetic field can be obtained using the usual Ampere's law 
    as $\bm j =4\pi \bm\nabla\times \bm B $. 
   
     To illustrate these results let us consider the particular case of
    spherically symmetric inhomogeneity modelled by  $\eta_2 (\bm r) = \eta_{20} e^{-r^2}$. 
   Magnetic  field and current patterns 
     produced by such defects are shown in Fig.(\ref{FigS7}). 
    One can see that the directions of fields and currents are opposite in $s+is$ and $s-is$ states. 
     The magnetic field distributions (left panels) are symmetric with respect to all crystal symmetries,
     that is rotations around the $z$ axis by
      an arbitrary angle as well as the $\pi$ rotation around the $x$ or $y$ axes. 
       The field pattern consists of the two clockwise and counter-clockwise parts    
      marked by red and blue arrows, respectively. This supercurrent (right panels in \ref{FigS7}) 
      has the same symmetry, therefore only the $z$-components of the current $j_z$ contribute to the 
      magnetic field generation. Currents with $j_z>(<)0$ produce clockwise (counter-clockwise) magnetic fields
   so that we show these currents by the red (blue) fonts, respectively in Fig.(\ref{FigS7}). 
          
          The overall sign of the magnetic field in Eqs.(\ref{Eq:mfGL1x},\ref{Eq:mfGL1y}), 
          is determined by the difference between the two gradient coefficients 
          $k_{12}^z-k_{12}^{xy} = K_{1}^z - K_{1}^{xy}  - K_{2}^z + K_{2}^{xy} $.
          Depending on whether this combination is positive or negative the magnetic field
          created by the spherically symmetric impurity in the $s+is'$ state 
           goes clockwise or counter-clockwise at $z>0$ and vice versa at $z<0$.
           The situation shown in Fig.(\ref{FigS7}) corresponds to $(k_{12}^z-k_{12}^{xy})>0$. 
          
%

 The absence of magnetic field $z$-component in 
 Fig.(\ref{FigS7}) results from the linear approximation that we used for obtaining the 3D field distribution. 
 We can check the validity of this assumption by comparing the magnitudes of $\bm B\parallel c$
 and $\bm B\parallel ab$ components of the spontaneous magnetic field
 by   solving the full GL model (\cite{Vadimov2018s}) for the effectively 2D  
 systems. For that purpose we assume the following distributions of the pairing coefficient
%
%
 \begin{align} \label{Eq:gaussianXY}
 \eta_2(x,y) =1 + \delta\eta_2 e^{-(x^2+y^2)/2} & & \textnormal{for $ab$-defect  }, 
 \\
 \label{Eq:gaussianXZ}
 \eta_2 (x,z)= 1 + \delta\eta_2 e^{-(x^2+z^2)/2} & & \textnormal{for $ca$-defect   } ,
 \end{align}
 where the length is normalized to the the superconducting coherence length $\xi$. The first 
 Eq.\ (\ref{Eq:gaussianXY}) describes the  inhomogeneity in the crystal $ab$-plane, where one has the C$_4$ rotation symmetry. 
 The second 
 Eq.\ (\ref{Eq:gaussianXZ}) describes the inhomogeneity in the $ac$-plane.

\begin{figure}[t]
\includegraphics[width=20pc,clip]{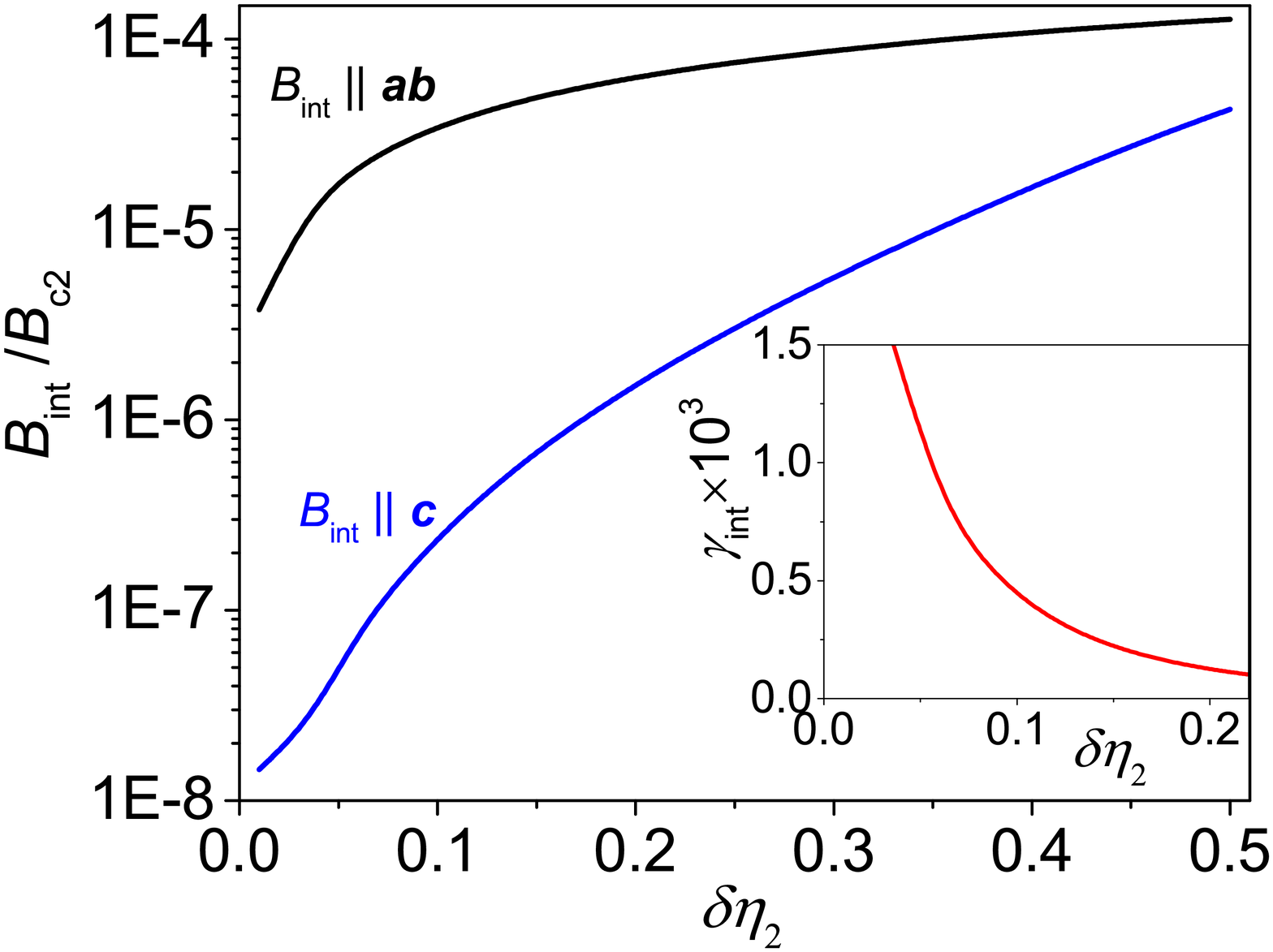}
\caption{Two orthogonal components of the normalized internal spontaneous fields $B_{\rm int}/B_{\rm c2}$ depending on a strength of the variation of the superconducting coupling constant $\delta\eta_2$ due to an inhomogeneity. Inset: $\delta\eta_2$ dependence of the anisotropy ratio of the internal fields $\gamma_{\rm int}$.} 
\label{FigS6}
\end{figure}

 For small variations of $\eta_2$ one can assume 
$\lambda  =  \lambda_0 \pm a\delta\eta_2$, where $a \sim 1$ is some numerical constant. 
The $\delta\eta_2$ dependence of the magnitude of the spontaneous fields in the $ab$-plane
 $B_{{\rm int}}\parallel ab$ for $ac$-oriented defects and along the $c$-axis $B_{{\rm int}}\parallel c$ 
 for $ab$-oriented defects are shown in Fig.\ \ref{FigS6}. The field is given in units of the upper critical field, $B_{\rm c2}\parallel c \sim 100$ kG with an anisotropy ratio 
 $\gamma_{\rm Hc2} = \frac {B_{\rm c2}\parallel ab}{B_{\rm c2}\parallel c} \sim 3 $ for
  the doping range near $x \sim 0.8$.\cite{Liu2014s} The resulting anisotropy of the internal
   fields $\gamma_{\rm int} = \frac {B_{{\rm int}}\parallel ab}{B_{{\rm int}}\parallel c}$ is enhanced additionally by the anisotropy of $\gamma_{\rm Hc2}$ as shown in the inset of Fig.\ \ref{FigS6}. It is seen that $\gamma_{\rm int} \gtrsim 10^3$ for weak inhomogeneities with $\delta\eta_2 < 0.05$.

To estimate the average internal fields $<B_{\rm int}>$ we assumed a normal distribution of the $\delta\eta_2$ values within the sample volume. According to the main text $\delta\lambda\approx\delta\eta_2$ varies in the interval of $\sim \pm 5 \%$. This interval corresponds roughly to $95 \%$ of the sample volume according to the experimental error-bars. Then taking into account a quasilinear dependence of $B_{\rm int}$ on $\delta\eta_2$ below 0.1, we arrive at average spontaneous fields of $<B_{\rm int}> \sim 0.2$ Oe (with the maximum fields up to several Oe), which is in a surprisingly good agreement (taking into account used simplifications) with the average value estimated from the $\mu$SR data. We believe that our work will stimulate a development of more realistic and sophisticated models based on more experimental material parameters.
 
\vspace{0.2cm}


\begin{thebibliography}{100}
\bibitem{note_Fig1} The propeller-like pocket around the $X$-point and an additional $h$- pocket in the zone center is not shown for the sake of simplicity since according to the simplest scenario for the BTRS state, these pockets do not play a leading role in the superconducting pairing in the doping range where the BTRS state is observed.
\bibitem{footnote1} Different transverse and the longitudinal contributions to the depolarization rate for $P_{\rm \mu} \parallel a$ can be expected if the internal fields are aligned within the $ab$-plane. The transverse component enhances the depolarization rate similarly to the observations for $P_{\rm \mu} \parallel c$, whereas the longitudinal component of the spontaneous fields can decrease the depolarization rate due to small static internal fields related to the nuclear moments $\sigma/\gamma_{\rm n} \sim 1$ Oe.  Our theoretical analysis predicts spontaneous fields up to several Oe with the average value of about 0.2 Oe  (see Section 5 in the SM). The overall effect depends strongly on the details of the spontaneous field distribution, which impedes a quantitative analysis of the experimental data in the BTRS state for $P_{\rm \mu} \parallel a$.
\end{thebibliography}
\end{document}